\pdfoutput=1

\documentclass[10pt]{iopart}
\usepackage{iopams}

\usepackage{amsfonts,amssymb} 
\usepackage{bm}
\usepackage{graphicx}
\usepackage{enumerate}
\usepackage{hyperref} 

\newtheorem{theorem}{Theorem}[section]
\newtheorem{remark}{Remark}[section]

\newcommand{\R}{\mathbb{R}}
\newcommand{\C}{\mathbb{C}}
\newcommand{\Z}{\mathbb{Z}}

\newcommand{\B}{\mathcal{B}}
\newcommand{\brill}{{\mathcal{B}}}

\newcommand{\bk}{{\bf k}}

\newcommand{\bK}{{\bf K}}
\newcommand{\bKp}{{\bf K'}}
\newcommand{\bv}{{\bf v}}

\newcommand{\bx}{{\bf x}}

\newcommand{\by}{{\bf y}}

\newcommand{\vtilde}{{\bm{\mathfrak{v}}}}
\newcommand{\ktilde}{{\bm{\mathfrak{K}}}}

\newcommand{\kpar}{{k_{\parallel}}}
\newcommand{\kparpi}{{\kpar=2\pi/3}}
\newcommand{\kparv}{{\kpar=\bK\cdot\vtilde_1}}

\newcommand{\inner}[1]{\left\langle#1\right\rangle}
\newcommand{\D}{\partial}

\newcommand{\eps}{\varepsilon}
\newcommand{\nit}{\noindent}
\newcommand{\nn}{\nonumber}
\newcommand{\sgn}{sgn}

\newcommand{\thetasharp}{{\vartheta_{\sharp}}}

\newcommand{\sech}{\textrm{sech}}

\begin{document}

\title[Topologically protected and non-protected edge states in 2D honeycomb structures]
{Bifurcations of edge states -- topologically protected and non-protected -- in continuous 2D honeycomb structures}

\date{\today}

\author{C.~L. Fefferman$^1$, J.~P. Lee-Thorp$^2$ and M.~I. Weinstein$^3$}

\address{$^1$ Department of Mathematics, Princeton University, Princeton, NJ, USA}
\address{$^2$ Department of Applied Physics and Applied Mathematics, Columbia University, New York, NY, USA}
\address{$^3$ Department of Applied Physics and Applied Mathematics and Department of Mathematics, Columbia University, New York, NY, USA}

\ead{\mailto{cf@math.princeton.edu}, \mailto{jpl2154@columbia.edu}, \mailto{miw2103@columbia.edu}}

\begin{abstract}
Edge states are time-harmonic solutions to energy-conserving wave equations, which are propagating parallel to a line-defect or ``edge'' and are localized transverse to it. 
This paper summarizes and extends the authors' work on the bifurcation of topologically protected edge states in continuous two-dimensional honeycomb structures.  

We consider a family of Schr\"odinger Hamiltonians consisting of a bulk honeycomb potential and a perturbing edge potential. The edge potential interpolates between two different periodic structures via a domain wall. 
We begin by reviewing our recent bifurcation theory of edge states for continuous two-dimensional honeycomb structures (http://arxiv.org/abs/1506.06111). The topologically protected edge state bifurcation is seeded by the zero-energy eigenstate of a one-dimensional Dirac operator. We contrast these protected bifurcations with (more common) non-protected bifurcations from spectral band edges, which are induced  by bound states of an effective Schr\"odinger operator.   
 
Numerical simulations for honeycomb structures of varying contrasts and ``rational edges'' (zigzag, armchair and others), support the following scenario: (a) For low contrast, under a sign condition on a distinguished Fourier coefficient of the bulk honeycomb potential, there exist topologically protected edge states localized transverse to zigzag edges. Otherwise, and for general edges, we expect long lived {\it edge quasi-modes} which slowly leak energy into the bulk.  (b) For an arbitrary rational edge, there is a threshold in the medium-contrast (depending on the choice of edge) above which there exist topologically protected edge states. 
In the special case of the armchair edge, there are two families of protected edge states; for each parallel quasimomentum (the quantum number associated with translation invariance) there are edge states which propagate in opposite directions along the armchair edge. 
\end{abstract}

\submitto{\TDM}

\maketitle

\ioptwocol

\section{Introduction and Summary}\label{intro}

Edge states are time-harmonic solutions to energy-conserving wave equations, which are propagating parallel to a line-defect or ``edge'' and localized transverse to it. 
This paper summarizes and extends the authors' work on the bifurcation of topologically protected edge states in two-dimensional (2D) honeycomb structures. 

We introduce a rich class of Schr\"odinger Hamiltonians consisting of a bulk honeycomb potential and perturbing edge potential. The perturbed Hamiltonian interpolates between Hamiltonians for two different asymptotic periodic  structures  via a domain wall. Localization transverse to ``hard edges''  and domain-wall induced edges has been explored quantum, photonic, and more recently,  acoustic, elastic and mechanical one- and two-dimensional systems; see, for example, \cite{kane2005z,kane2005quantum,Su-Schrieffer-Heeger:79,HR:07,RH:08,malkova2009observation,Shvets-PTI:13, Kane-Lubensky:13, plotnik2013observation,rechtsman2013photonic,Shvets:14,FLW-PNAS:14,Sheng:15,MKW:15,PCV:15,ablowitz2015adiabatic,yang2015topological,susstrunk2015observation}. 

In Section  \ref{preliminaries} we present the necessary background on  honeycomb structures, their band structures and, in particular, their ``Dirac points''. These are quasimomentum/energy pairs whose dispersion surfaces touch conically, a 2D version of linear band crossings in one dimension (1D). For honeycomb structures, Dirac points occur at the vertices of the hexagonal Brillouin zone, which are high-symmetry quasimomenta.  Section \ref{edge_model} introduces a family of Schr\"odinger Hamiltonians, $H^{(\eps,\delta)}$,  consisting of a bulk honeycomb part, $H^{(\eps,0)}=-\Delta +\eps V(\bx)$, plus a perturbing edge potential,  $\delta\kappa(\delta\ktilde_2\cdot \bx)W(\bx)$; see \eref{schro-domain}. 
Here, $\eps$ measures the bulk medium-contrast and $\delta$ parametrizes the strength and spatial scale of the edge perturbation.

An edge state, which is localized transverse to a ``rational edge'', is a non-trivial solution of the eigenvalue problem, $H^{(\eps,\delta)}\Psi=E\Psi$, where $\Psi$ lies in a Hilbert space of functions  defined on an appropriate cylinder ($\R^2$ modulo the rational edge).
In Section \ref{thm-edge-states} we discuss our general bifurcation theory of topologically protected bifurcations of edge states. 
This bifurcation is governed by the zero-energy eigenstate of an effective 1D Dirac operator. Theorem \ref{general-conditions}, proved in \cite{FLW-JAMS:15},  provides general conditions for the existence of an edge state, which is localized transverse to a rational edge.
A key role is played a spectral property of the bulk (unperturbed) honeycomb Hamiltonian, which we  call the {\it spectral no-fold condition}. Section \ref{zigzag-summary} discusses the application of Theorem \ref{general-conditions} to topologically protected bifurcations of edge states transverse to a zigzag edge.

In Section \ref{unprotected} we compare such topologically protected bifurcations with the more typical bifurcations of localized states from spectral band edges. The latter are governed by the localized eigenstates of an effective Schr\"odinger operator and are not topologically protected. Bifurcations from spectral band edges are sometimes called ``Tamm states'' while those which occur at linear band crossings  are sometimes called ``Shockley states'' \cite{Tamm:32,Shockley:39,malkova2009observation}.
Finally, in Section \ref{numerics_general_edges} we present and interpret numerical simulations for a variety of rational edges. The specific honeycomb potential and domain wall/edge perturbation is displayed in \eref{VW-numerics}. These include zigzag, armchair and other edges. 
 These investigations support the following scenario: 
\begin{itemize}
\item[(a)] For low contrast, under a sign condition on a distinguished Fourier coefficient of the bulk honeycomb potential, there exist topologically protected edge states localized transverse to zigzag edges. Otherwise, and for general edges, we expect long lived {\it edge quasi-modes} which slowly leak energy into the bulk. 
\item[(b)] For an arbitrary rational edge, there is a threshold, $\eps_0\ge0$, in the medium-contrast (depending on the choice of edge) such that for $\eps>\eps_0$, there exist topologically protected edge states. In particular, $\eps_0({\rm zigzag})=0$ and $\eps_0({\rm armchair})>0$.
\item[(c)]
In the special case of the armchair edge, for $\eps>\eps_0$ there are two families of protected edge states; for each parallel quasimomentum (the quantum number associated with translation invariance) there are edge states which propagate in opposite directions along the armchair edge. 
\end{itemize}

%
%
%
%
%
A complete analytical understanding of these observations is work in progress.

\section{Honeycomb potentials and Dirac points}\label{preliminaries}

\subsection{Honeycomb potentials}\label{honeycomb_potentials}

\begin{figure}
\centering
\includegraphics[width=\columnwidth]{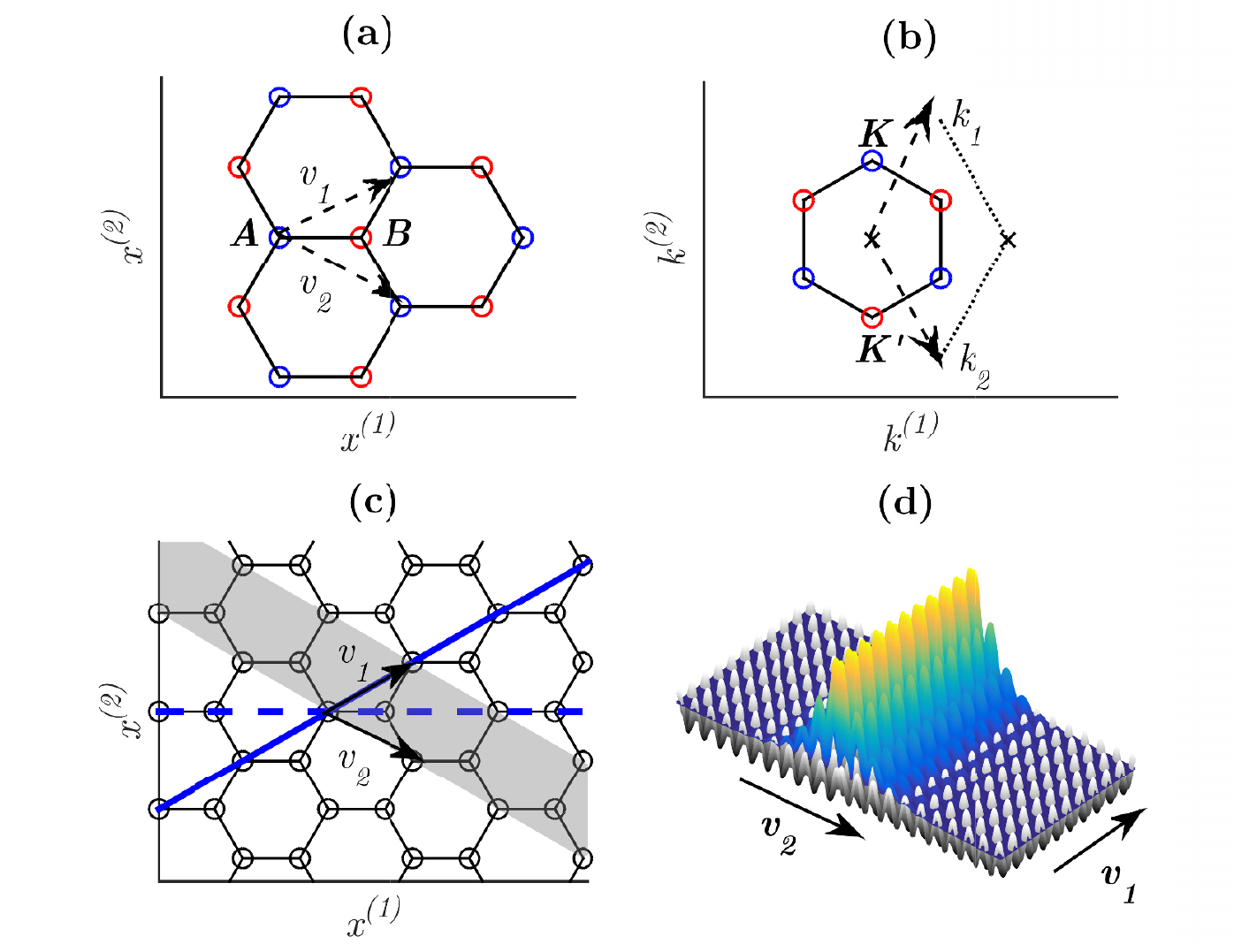}
\caption{
{\bf (a)}: ${\bf A}=(0,0)$, ${\bf B}=(\frac{1}{\sqrt3},0)$. 
The honeycomb structure, ${\bf H}$,  is the union of two sub-lattices  $\Lambda_{\bf A}={\bf A}+\Lambda_h$ (blue) 
and $\Lambda_{\bf B}={\bf B}+\Lambda_h$ (red). The lattice vectors  $\{\bv_1,\bv_2\}$ generate $\Lambda_h$.
{\bf (b)}:
Brillouin zone, $\brill_h$, and dual basis $\{\bk_1,\bk_2\}$. $\bK$ and $\bK'$ are labeled. 
{\bf (c)}:
Zigzag edge (solid line), $\R\bv_1 = \{\bx : \bk_2\cdot\bx=0\}$, armchair edge (dashed line), $\R\left(\bv_1+\bv_2\right) = \{\bx : (\bk_1-\bk_2)\cdot\bx=0\}$, and a fundamental domain for the cylinder for the zigzag edge (gray area), $\Sigma_{\rm ZZ}$.
{\bf (d)}: Schematic of 
edge state, localized transverse to the zigzag edge ($\R\bv_1$).
\label{fig_honeycomb_summary}
}
\end{figure}

Introduce the basis vectors: ${\bf v}_1=(\frac{\sqrt{3}}{2} , \frac{1}{2})^T$,  ${\bf v}_2=(\frac{\sqrt{3}}{2} , -\frac{1}{2})^T$ and
dual basis vectors: $\bk_1=  q( \frac{1}{2} , \frac{\sqrt{3}}{2} )^T$, $\bk_2=  q( \frac{1}{2} , -\frac{\sqrt{3}}{2} )^T$,  where $ q\equiv \frac{4\pi}{\sqrt{3}}$, which satisfy the relations  $\bk_l\cdot\bv_m=2\pi \delta_{lm},\ l,m=1,2$.
Let $\Lambda_h = \Z\bv_1\oplus \Z\bv_2$ denote the regular (equilateral) triangular lattice and $\Lambda_h^* = \Z\bk_1\oplus \Z\bk_2$,  the associated dual lattice. 
The honeycomb structure, ${\bf H}$, is the union of the two interpenetrating triangular lattices: ${\bf A} + \Lambda_h$ and ${\bf B} + \Lambda_h$, where ${\bf A}=(0,0)^T$ and ${\bf B}=(\frac{1}{\sqrt{3}},0)^T$; see Figure \ref{fig_honeycomb_summary}(a).
$\mathcal{B}_h$ denotes the Brillouin zone, the choice of fundamental cell in quasi-momentum space is displayed in Figure \ref{fig_honeycomb_summary}(b).

A \emph{honeycomb lattice potential}, $V(\bx)$,  is a real-valued,  smooth function, which is $\Lambda_h-$ periodic and, relative to some origin of coordinates, is  inversion symmetric  (even) and invariant under a $2\pi/3$ rotation. Specifically, if $R$ denotes the $2\times2$ rotation matrix by $2\pi/3$, we have 
$\mathcal{R}[V](\bx)\equiv V(R^*\bx)=V(\bx).$ 
  A natural choice of period cell is $\Omega_h$, the parallelogram in $\R^2$ spanned by $\{\bv_1, \bv_2\}$.
   
Consider the Hamiltonian for the unperturbed honeycomb structure:
\begin{equation}
H^{(\eps,0)}  = -\Delta + \eps V(\bx).        \label{H0} 
\end{equation}
The \emph{band structure} of the $\Lambda_h-$ periodic Schr\"odinger operator, $H^{(\eps,0)}$, is obtained from the
family of eigenvalue problems, parametrized by $\bk\in\mathcal{B}_h$: 
 $(H^{(\eps,0)}-E)\Psi=0,\ \Psi(\bx+\bv)=e^{i\bk\cdot\bv}\Psi(\bx),\ \ \bx\in\R^2,\ \bv\in\Lambda_h$.
We denote by $L^2_\bk$ the space of functions $f\in L^2_{\rm loc}$ satisfying the $\bk-$ pseudo-periodic boundary condition $f(\bx+\bv)=e^{i\bk\cdot\bv}f(\bx)$.
For $f$ and $g$ in $L^2_\bk$, $\overline{f}g$ is in $L^2(\R^2/\Lambda)$ and we define their inner product by
$\inner{f,g}_{L^2_\bk} = \int_\Omega \overline{f(\bx)} g(\bx) d\bx .$
 Equivalently,   $\psi(\bx)=e^{-i\bk\cdot\bx}\Psi(\bx)$ satisfies the periodic eigenvalue problem:
$ \left(H^{(\eps,0)}(\bk)-E(\bk)\right)\psi=0$ and $\psi(\bx+\bv)=\psi(\bx)$ for all $\bx\in\R^2$ and 
$\bv\in\Lambda_h$, where   
\begin{equation*}
H^{(\eps,0)}(\bk)=-(\nabla+i\bk)^2+ \eps V(\bx).
\label{H0k}\end{equation*}
 For each $\bk\in\B_h$, the spectrum is real and consists of discrete eigenvalues  $E_b(\bk),\ b\ge1, $ where $E_b(\bk)\le E_{b+1}(\bk)$. The graphs of  $\bk\mapsto E_b(\bk)\in\R$ are called the dispersion surfaces of $H^{(\eps,0)}$. The collection of these surfaces constitutes the band structure of $H^{(\eps,0)}$. As $\bk$ varies over $\mathcal{B}_h$, each map $\bk\to E_b(\bk)$ is Lipschitz continuous and sweeps out a  closed interval in $\R$. The union of these intervals is the $L^2(\R^2)-$ spectrum of $H^{(\eps,0)}$.
More detail on the general theory of periodic operators and  the specific context of honeycomb structures appears in \cite{RS4, Eastham:74, kuchment2012floquet} and \cite{FW:12,FLW-JAMS:15}, respectively.

\subsection{Dirac points}\label{dirac_points}

Let $V$ denote a honeycomb lattice potential. \emph{Dirac points}  of $H^{(\eps,0)}$ are quasimomentum/energy pairs, $(\bK_\star,E_\star)$,  in the band structure of $H^{(\eps,0)}$ at which neighboring dispersion surfaces touch conically at a point \cite{RMP-Graphene:09,Katsnelson:12,FW:12}. In particular, there exists a $b_\star\ge1$  such that:
\begin{enumerate}
\item $E_\star=E_{b_\star}(\bK_\star)=E_{b_\star+1}(\bK_\star)$ is a two-fold degenerate $L^2_{\bK_\star}-$ eigenvalue of  $H^{(\eps,0)}$.
\item $\textrm{Nullspace}(H^{(\eps,0)}-E_\star I) = 
{\rm span}\{ \Phi_1(\bx) , \Phi_2(\bx)\}$, 
where $\Phi_1\in L^2_{\bK_\star,\tau} = L^2_{\bK_\star}\cap\{f: \mathcal{R}f=\tau f\}$ and 
$\Phi_2(\bx) = \overline{\Phi_1(-\bx)}\in L^2_{\bK_\star,\bar\tau}=L^2_{\bK_\star}\cap\{f:\mathcal{R}f=\overline{\tau}f\}$,
and $\inner{\Phi_a, \Phi_b}_{L^2_{\bK_\star}(\Omega)} = \delta_{ab}$, $a,b=1,2$. Here, $\tau=\exp(2\pi i/3)$. We note that $1, \tau$ and $\overline\tau$ are eigenvalues of the rotation matrix, $R$. 
%
%
%
%
\item There exist $L^2_{\bk}-$ Floquet-Bloch eigenpairs 
$\bk\mapsto (\Phi_{b_\star}(\bx;\bk),E_{b_\star}(\bk))$ and $\bk\mapsto (\Phi_{b_\star+1}(\bx;\bk),E_{b_\star+1}(\bk))$, defined in a neighborhood of $\bK_\star$, such that
\begin{equation*}
\eqalign{
E_{b_\star+1}(\bk)-E_\star \approx + |\lambda_\sharp|
\left| \bk-\bK_\star \right| ,\\
E_{b_\star}(\bk)-E_\star \approx - |\lambda_\sharp|
\left| \bk-\bK_\star \right|. }
\end{equation*}
where $|\lambda_\sharp|\  = 2 |\inner{\Phi_2, {\zeta}\cdot\nabla\Phi_1}_{L^2_{\bK_\star}}| \neq 0$ for any $\zeta=(\zeta_1,\zeta_2)\in\C^2$, such that $\|\zeta\|_{\C^2}=1$. 
\end{enumerate}

\begin{figure}
\centering
\includegraphics[width=\columnwidth]{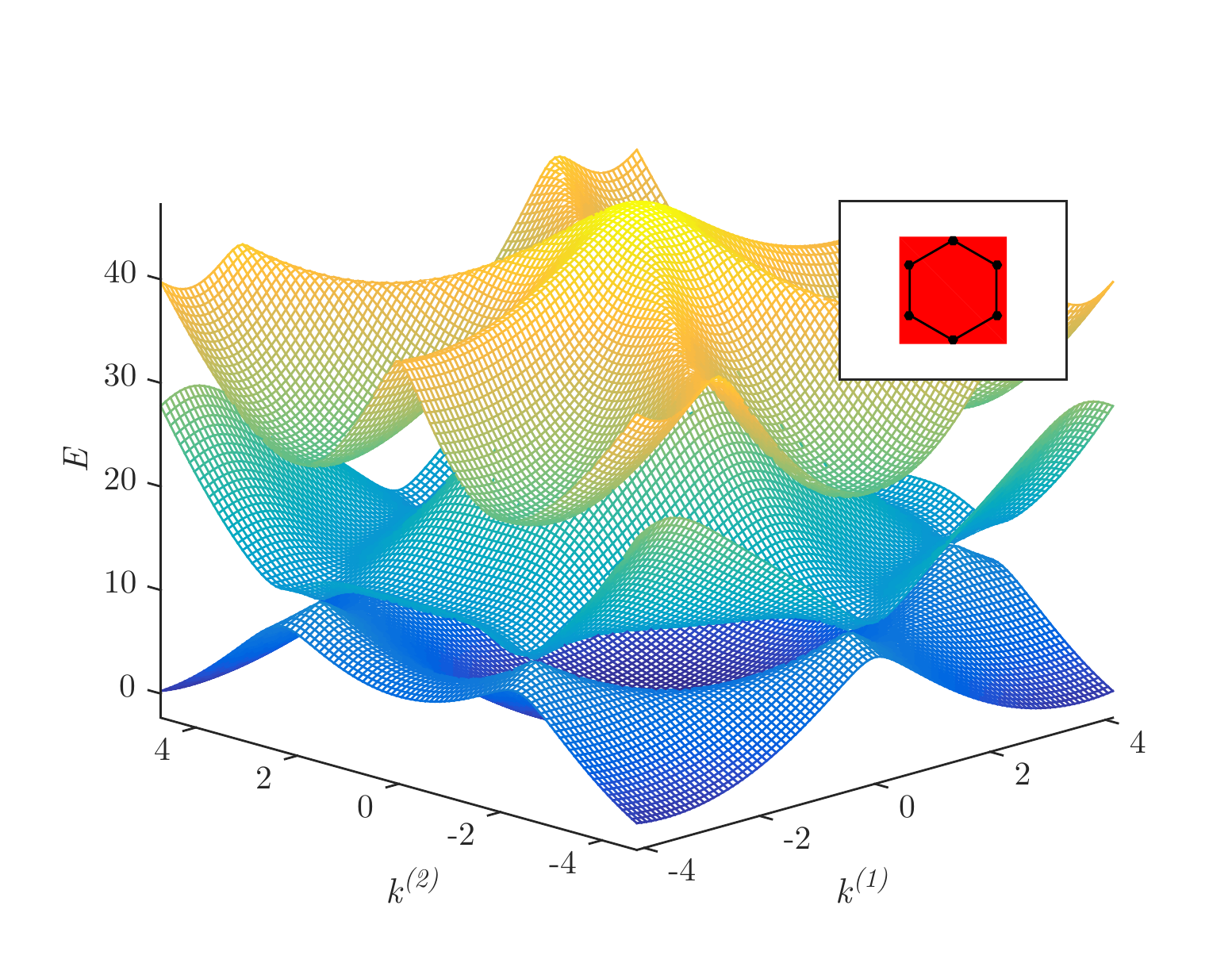}
\caption{
Lowest three dispersion surfaces $\bk\equiv(k^{(1)},k^{(2)})\in\mathcal{B}_h\mapsto E(\bk)$ of the band structure of  $H^{(10,0)}\equiv -\Delta + 10 V(\bx)$, where 
$V$ is a honeycomb potential: $V(\bx) = \left(\cos(\bk_1 \cdot\bx)+\cos(\bk_2 \cdot\bx)+\cos((\bk_1+\bk_2)\cdot\bx)\right)$.
Dirac points occur at the conical intersection of the lower two dispersion surfaces, at the six vertices of the Brillouin zone, $\mathcal{B}_h$. 
\label{fig_E_vs_k_three_surfaces}
}
\end{figure}

\noindent In \cite{FW:12} (see also \cite{FLW-MAMS:15}, Appendix D), Fefferman and Weinstein proved  the following:
   
\begin{theorem}\label{generic-dirac-points}
(a) For all $\eps$ real outside a possible discrete set, the Hamiltonian $H^{(\eps,0)}=-\Delta+\eps V$ has Dirac points occurring at $(\bK_\star,E_\star)$, where $\bK_\star$ may be any of the six vertices of the  Brillouin zone, $\mathcal{B}_h$.
(b) If $0<|\eps|<\eps_0$ is sufficiently small, then there are two cases, which are delineated by the sign of the distinguished Fourier coefficient, $\eps V_{1,1}$, of $\eps V(\bx)$. Here, 
\begin{equation}
\qquad  V_{1,1} \equiv 
\frac{1}{|\Omega_h|} \int_{\Omega_h} e^{-i(\bk_1+\bk_2)\cdot\by} V(\by) d\by,
\label{V11eq0-intro}
\end{equation}
is assumed to be non-zero. If $\eps V_{1,1}>0$, then $b_\star=1$; Dirac points occur at the intersection of the first and second dispersion surfaces (see Figure \ref{fig_E_vs_k_three_surfaces}).  If $\eps V_{1,1}<0$, then $b_\star=2$;
they occur at the intersection of the second and third dispersion surfaces.  
 \end{theorem} 
 The two cases in part (b) of Theorem \ref{generic-dirac-points} are illustrated 
 in Figure \ref{fig_spectral_nofold} (top row, left and center plots).

The quasimomenta of Dirac points partition into two equivalence classes; the $\bK-$ points ($\bK=\frac{1}{3}\left(\bk_1-\bk_2\right)$) consisting of $\bK, R\bK$ and $R^2\bK$, and  $\bK'-$ points ($\bKp=-\bK$) consisting of $\bK'=-\bK, R\bK'$ and $R^2\bK'$; see Figure \ref{fig_honeycomb_summary}(b).
By symmetries, the local character near one Dirac point determines the local character near the others.
 
The time evolution of a  wave packet, with data spectrally localized near a Dirac point, is governed by a massless 2D Dirac system \cite{FW:14}.

\begin{remark}\label{symmetry_breaking}
It is shown in \cite{FW:12} that a $\Lambda_h-$ periodic perturbation of $V(\bx)$, which breaks inversion or time-reversal symmetry, lifts the eigenvalue degeneracy; a (local) gap is opened about the Dirac points and the perturbed dispersion surfaces are locally smooth. The edge potential we introduce opens such a gap, as illustrated in the bottom row of Figure \ref{fig_spectral_nofold}.
\end{remark}
  
\section{Honeycomb structure with an edge and the edge state eigenvalue problem}\label{edge_model}
 
We follow the setup introduced in  \cite{FLW-JAMS:15}.  Recall (Section \ref{honeycomb_potentials}) the spanning vectors of the equilateral triangular lattice, $\bv_1$ and $\bv_2$. Given  a pair of  integers $a_1, b_1$, which are relatively prime, let $\vtilde_1=a_1\bv_1+b_1\bv_2$. We call the line $\R\vtilde_1$ the $\vtilde_1-$ edge. Since $a_1, b_1$ are relatively prime,  there exists a second pair of  relatively prime integers: $a_2,b_2$ such that $a_1b_2-a_2b_1=1$.  Set $\vtilde_2 = a_2 \bv_1 + b_2 \bv_2$. 

It follows that $\Z\vtilde_1\oplus\Z\vtilde_2=\Z\bv_1\oplus\Z\bv_2=\Lambda_h$.
Since $a_1b_2-a_2b_1=1$, we have dual lattice vectors $\ktilde_1, \ktilde_2\in\Lambda_h^*$, given by
$ \ktilde_1=b_2\bk_1-a_2\bk_2$ and $ \ktilde_2=-b_1\bk_1+a_1\bk_2$, which satisfy
$\ktilde_\ell \cdot \vtilde_{\ell'} = 2\pi \delta_{\ell, \ell'}$, $1\leq \ell, \ell' \leq 2$.
Note that $\Z\ktilde_1\oplus\Z\ktilde_2=\Z\bk_1\oplus\Z\bk_2=\Lambda^*_h$.  
We denote by $\Omega$ the period cell given by the parallelogram  spanned by $\{\vtilde_1, \vtilde_2\}$.
The choice $\vtilde_1=\bv_1$ (or equivalently $\bv_2$) generates a \emph{zigzag edge} and the choice $\vtilde_1=\bv_1+\bv_2$ generates the \emph{armchair edge}; see Figure \ref{fig_honeycomb_summary}(c). 
 
Introduce the  family of Hamiltonians, depending on the real parameters $\eps$ and $\delta$:
\begin{equation}
H^{(\eps,\delta)} \equiv -\Delta + \eps V(\bx) + \delta\kappa(\delta\ktilde_2\cdot \bx)W(\bx). \label{schro-domain} 
\end{equation}
$H^{(\eps,0)}=-\Delta +\eps V(\bx)$ is the Hamiltonian for the unperturbed (bulk) honeycomb structure, introduced in \eref{H0} and discussed in Theorem \ref{generic-dirac-points}.
Here, $\delta$ will be taken to be sufficiently small, and $W(\bx)$ is $\Lambda_h-$ periodic
and odd. The function $\kappa(\zeta)$ defines a \emph{domain wall}. We choose $\kappa$ to be sufficiently smooth and to satisfy $\kappa(0)=0$ and    $\kappa(\zeta)\to\pm\kappa_\infty\ne0$
as $\zeta\to\pm\infty$, \emph{e.g.} $\kappa(\zeta)=\tanh(\zeta)$. Without loss of generality, we assume $\kappa_\infty>0$. 

Note that $H^{(\eps,\delta)}$ is invariant under translations  parallel to the  $\vtilde_1-$ edge, $\bx\mapsto\bx+\vtilde_1$, and hence there is a well-defined \emph{parallel quasimomentum} (good quantum number), denoted $\kpar$.  Furthermore, $H^{(\eps,\delta)}$ transitions adiabatically between the  asymptotic Hamiltonian $H_-^{(\eps,\delta)}=H^{(\eps,0)} - \delta\kappa_\infty W(\bx)$ as $\ktilde_2\cdot\bx\to-\infty$ to the asymptotic Hamiltonian $H_+^{(\eps,\delta)}=H^{(\eps,0)} + \delta\kappa_\infty W(\bx)$ as $\ktilde_2\cdot\bx\to\infty$. 
The  domain wall modulation of $W(\bx)$ realizes a phase-defect across the edge $\R\vtilde_1$.
 A variant of this construction was used in \cite{FLW-PNAS:14,FLW-MAMS:15} to interpolate between different asymptotic 1D dimer periodic potentials. 
 
We seek  \emph{$\vtilde_1-$ edge states} of $H^{(\eps,\delta)}$, which are spectrally localized near the Dirac point, $(\bK_\star,E_\star)$. 
These are non-trivial solutions $\Psi$, with energies $E\approx E_\star$,  of the \emph{$\kpar-$ edge state eigenvalue problem (EVP):}
\begin{eqnarray}
H^{(\eps,\delta)}\Psi &= E\Psi,\label{edge-evp}\\
\Psi(\bx+\vtilde_1) &= e^{i\kpar}\Psi(\bx) , \label{edge-bc1}\\
|\Psi(\bx)| \to \ &0\ {\rm as} \ |\ktilde_2\cdot\bx|\to\infty , \label{edge-bc2}
\end{eqnarray}
for $\kpar\approx\bK_\star\cdot\vtilde_1$. The boundary conditions \eref{edge-bc1} and \eref{edge-bc2} imply, respectively, propagation parallel to, and localization transverse to, the edge $\R\vtilde_1$.

We next formulate the edge state eigenvalue problem \eref{edge-evp}-\eref{edge-bc2} in an appropriate Hilbert space. Introduce the cylinder $\Sigma\equiv \R^2/ \Z\vtilde_1$. 
Denote by  $H^s(\Sigma),\ s\ge0$, the  Sobolev spaces of functions defined on $\Sigma$. The pseudo-periodicity and decay conditions \eref{edge-bc1}-\eref{edge-bc2} are encoded by requiring $ \Psi \in H^s_\kpar(\Sigma)=H^s_\kpar$, for some $s\ge0$,  where
\[H^s_\kpar\equiv \ \left\{f : f(\bx)e^{-i\frac{\kpar}{2\pi}\ktilde_1\cdot\bx}\in H^s(\Sigma) \right\}.\]
We then formulate the EVP \eref{edge-evp}-\eref{edge-bc2} as:
\begin{equation}
H^{(\eps,\delta)}\Psi = E\Psi, \quad \Psi\in H^2_{\kpar}(\Sigma).
\label{EVP}
\end{equation}

\subsection{The spectral no-fold condition}\label{no-fold}

Now suppose $H^{(\eps,0)}$  has a Dirac point at $(\bK_\star,E_\star)$, {\it i.e.} $\eps$ is generic 
(not necessarily small) in the sense of Theorem \ref{generic-dirac-points}. 
  (Recall, from Section \ref{preliminaries} that all vertex quasimomenta of the hexagon, $\mathcal{B}_h$, have Dirac points with energy $E_\star$.)
While $H^{(\eps,0)}$ is inversion symmetric (invariant under $\bx\mapsto-\bx$),  $H^{(\eps,\delta)}_\pm$ is not.  Therefore, by Remark \ref{symmetry_breaking},
for $\delta\ne0$,  $H^{(\eps,\delta)}_\pm$ does not have Dirac points; its dispersion surfaces are locally smooth and for quasimomenta $\bk$ such that $|\bk-\bK_\star|$ is sufficiently small, there is an open neighborhood of $E=E_\star$ not contained in the  $L^2(\R^2/\Lambda_h)-$ spectrum of $H^{(\eps,\delta)}_\pm(\bk)$.  
  If there is a (real) open neighborhood of $E=E_\star$, not contained in the  spectrum of $H^{(\eps,\delta)}_\pm(\bk)$ for \emph{all} $\bk\in\B_h$, then 
$H_\pm^{(\eps,\delta)}$ is said to have a  (global) omni-directional spectral gap about $E=E_\star$.
But, the  ``spectral gap'' about $E=E_\star$, created for $\delta\ne0$ and small, may  only be local about $\bK_\star$. What is central however to the existence of $\vtilde_1-$ edge states is that along the dispersion surface slice, dual to the $\vtilde_1-$ edge, $H^{(\eps,0)}$ satisfy a \emph{spectral no-fold condition}. We now explain this condition, without going into all technical detail. 

\begin{figure}
\centering
\includegraphics[width=\columnwidth]{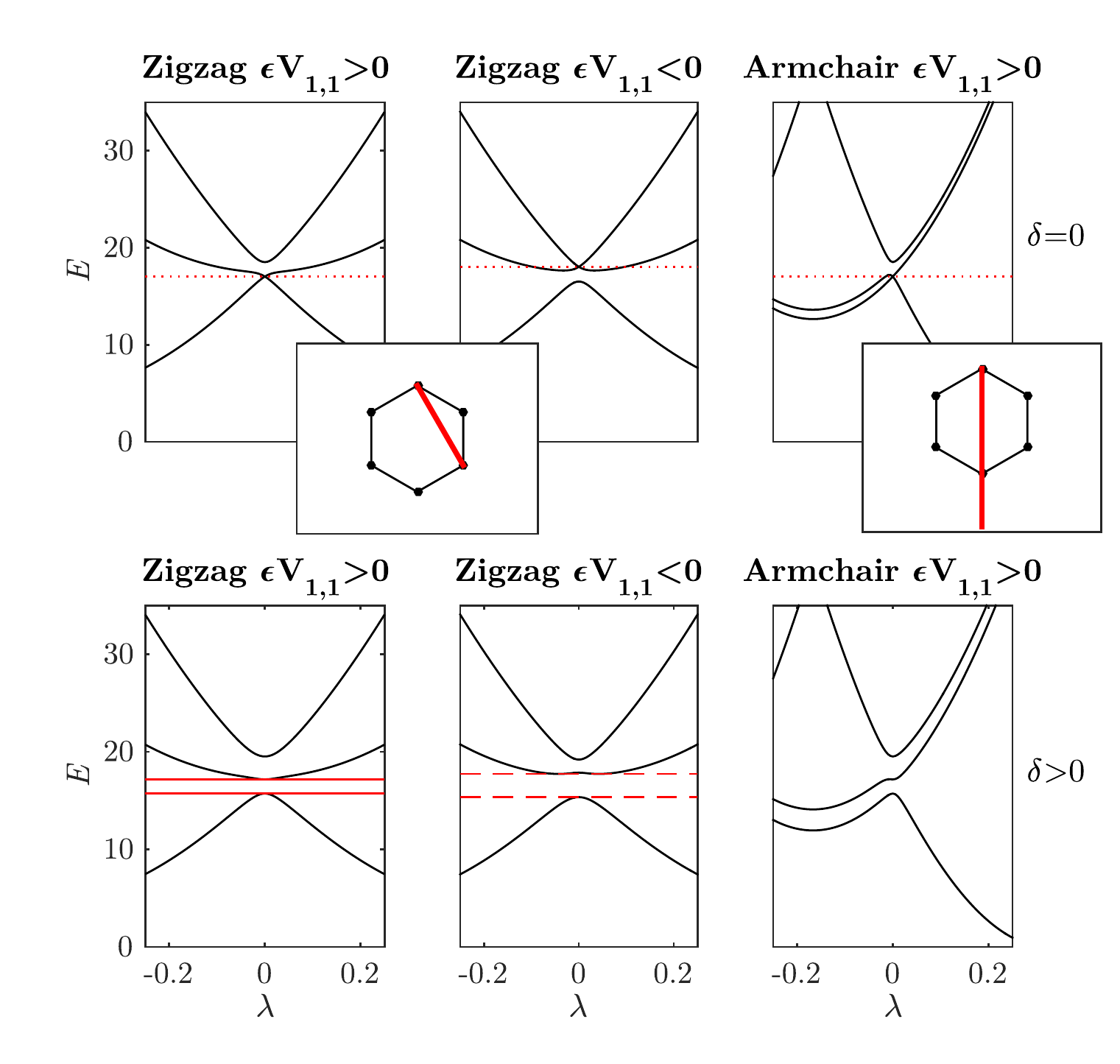}
\caption{
 Band dispersion slices along the quasimomentum segment:  $\bK+\lambda\ktilde_2,\ |\lambda|\le1/4$.
 In the  {\bf first row} ($\delta=0$), we consider whether the spectral no-fold condition holds at the Dirac point $(\bK,E_\star)$.
Energy level $E=E_\star$ is indicated with the dotted line.
The spectral no-fold condition holds along the zigzag slice for $\eps V_{1,1}>0$ and there is a topologically protected branch of edge states, but fails for $\eps V_{1,1}<0$ and along the armchair slice.
{\bf Second row} ($\delta>0$) of plots illustrates that the spectral no-fold condition controls whether a full directional spectral gap opens when breaking inversion symmetry.
 Insets indicate zigzag and armchair quasimomentum segments (1D Brillouin zones) parametrized by $\lambda$, for $0\leq\lambda\leq1$.
 \label{fig_spectral_nofold}
 }
\end{figure}

Let $(\bK_\star, E_\star)$ denote a Dirac point of $H^{(\eps,0)}$ in the sense of Section 
 \ref{generic-dirac-points}, where $E_\star=E_{b_\star}(\bK_\star)=E_{b_\star+1}(\bK_\star)$. Consider the {\it dual slice} associated with the $\vtilde_1-$ edge, {\it i.e.} the union of graphs of the  functions  $\lambda\in(-1/2,1/2]\mapsto E_b(\bK_\star +\lambda\ktilde_2)$, $b\ge1$. The values swept out constitute the $L^2_{\kpar=\bK_\star\cdot\vtilde_1}-$ spectrum. Within this  slice, the graphs of 
  $\lambda\mapsto E_{b_\star}(\bK_\star +\lambda\ktilde_2)$ and
  $\lambda\mapsto E_{b_\star+1}(\bK_\star +\lambda\ktilde_2)$  may intersect at either one or two independent Dirac points, {\it i.e.} points in the lattices $\bK+\Lambda_h^*$ or in $\bK'+\Lambda_h^*$, occurring at distinct values of $\lambda\in(-1/2,1/2]$.
({\it E.g.} for a zigzag edge there is one value, namely $\widehat{\lambda}_1=0$, and for an armchair edge there are two, namely $\widehat{\lambda}_1=0$ and $\widehat{\lambda}_2=-1/3$.) The preceding discussion implies that a  $L^2_{\kpar=\bK_\star\cdot\vtilde_1}-$ spectral gap of $H^{(\eps,\delta)}_\pm$ is opened \underline{locally} near each Dirac point, {\it i.e.} for all $\lambda$ near $\widehat\lambda_j$, by the domain wall/edge potential for $\delta\ne0$ and small; see Figure \ref{fig_spectral_nofold}.

\emph{ 
We say that the spectral no-fold condition is satisfied for the $\vtilde_1-$ edge if, for $\delta\ne0$, the $L^2_{\kpar=\bK_\star\cdot\vtilde_1}-$ spectrum of $H^{(\eps,\delta)}_\pm$ has a full spectral gap. That is, the $L^2_{\kpar=\bK_\star\cdot\vtilde_1}-$ spectrum of $H_\pm^{(\eps,\delta)}$ does not intersect some open interval containing $E_\star$.}
 
We note that this formulation of the no-fold condition is more general than that introduced in \cite{FLW-JAMS:15}. The role of the dual slice can be understood by recognizing that any function which satisfies the edge state boundary conditions, \eref{edge-bc1} and \eref{edge-bc2}, relative to the $\vtilde_1-$ edge is a superposition of Floquet-Bloch modes parametrized by the dual slice; see \cite{FLW-JAMS:15}. 

In the bottom panels of Figure \ref{fig_spectral_nofold}, we display examples of slices through the first three dispersion surfaces of $H^{(\eps,\delta)}$, where $\delta\ne0$. The middle and rightmost of these three plots illustrate that a dispersion surface may ``fold-over'' outside a neighborhood of the Dirac point, filling out all energies in a neighborhood of $E_\star$.

The bottom row of Figure \ref{fig_armchair_21_nofold} shows that for $\eps=10$ ($\eps$, large and positive) the spectral no-fold condition, as stated above, holds.
The bottom left panel of Figure \ref{fig_armchair_21_nofold} corresponds to the armchair slice where, for $\lambda\in(-1/2,1/2]$, the curves $\lambda\mapsto E_1(\bK_\star +\lambda\ktilde_2)$ and $\lambda\mapsto E_2(\bK_\star +\lambda\ktilde_2)$ intersect at two independent quasimomenta $\bK\in \bK+\Lambda_h^*$ and $\bK'\in \bK'+\Lambda_h^*$. The bottom right panel of Figure \ref{fig_armchair_21_nofold} corresponds to the $(2,1)$- slice ($\vtilde_1=2\bv_1+\bv_2$) where, for $\lambda\in(-1/2,1/2]$, the curves  $\lambda\mapsto E_1(\bK_\star +\lambda\ktilde_2)$ and $\lambda\mapsto E_2(\bK_\star +\lambda\ktilde_2)$ intersect at points in $\bK+\Lambda_h^*$.

\section{Topologically protected edge states}\label{thm-edge-states} 

\subsection{General conditions for  topologically protected bifurcations of edge states for rational edges}\label{general_conditions}

In this section we review conditions on the Hamiltonian, $H^{(\eps,\delta)}$, defined in \eref{schro-domain}, guaranteeing  the existence of topologically protected states for an arbitrary $\vtilde_1-$ edge. These results are proved in  \cite{FLW-JAMS:15}  (Theorem 7.3 and Corollary 7.4).

Let  $V$ be a honeycomb lattice potential (see Section \ref{honeycomb_potentials}) and $W$ be real-valued, $\Lambda_h-$ periodic and odd. 
Assume that $H^{(\eps,0)}=-\Delta+\eps V$ has a Dirac point
$(\bK_\star,E_\star)$, where $\bK_\star$ is a vertex of $\mathcal{B}_h$ as defined in Section \ref{dirac_points}.
Assume further that $H^{(\eps,0)}$ satisfies the {\it spectral no-fold condition} for the $\vtilde_1-$ edge, as stated in \cite{FLW-MAMS:15}. 
That is, we assume that the quasi-momentum slice through only intersects one independent Dirac point;
see the discussion of Section \ref{no-fold}.
\footnote{An article in which we extend this theorem to the situation where both classes of Dirac points lie in the dual slice is in preparation.}
Finally, we assume the non-degeneracy condition: 
\begin{equation}
\thetasharp\equiv\inner{\Phi_1,W\Phi_1}_{L_{\bK_\star}^2} \ne 0,
\label{theta-sharp}
\end{equation}
 where $\Phi_1(\bx)$ is the $L^2_{\bK_\star}-$ eigenfunction of $H^{(\eps,0)}$ associated with the quasimomentum $\bK_\star$ introduced in Section \ref{dirac_points}.

\begin{theorem}\label{general-conditions}

Under the above hypotheses:
\begin{enumerate}
\item There exist topologically protected $\vtilde_1-$ edge states with $\kpar=\bK_\star\cdot\vtilde_1$, constructed as a bifurcation curve of non-trivial  eigenpairs $\delta\mapsto (\Psi^\delta, E^\delta)$ of \eref{EVP}, defined for all $|\delta|$ sufficiently small. This branch of non-trivial states bifurcates from the trivial solution branch $E\mapsto(\Psi\equiv0,E)$ at $E=E_\star$. 
The edge state, $\Psi^\delta(x)$, is well-approximated (in $H_{\kparv}^2$) by a slowly varying and spatially decaying modulation of the degenerate nullspace of $H^{(\eps,0)}-E_\star$:
\begin{eqnarray*}
  \Psi^\delta(\bx) &\approx \alpha_{\star,+}(\delta\ktilde_2\cdot\bx)\Phi_+(\bx) + \alpha_{\star,-}(\delta\ktilde_2\cdot\bx)\Phi_-(\bx) , \\ 
  E^\delta &= E_\star + \mathcal{O}(\delta^2),\ \ 0<|\delta|\ll1,
\end{eqnarray*} 
where $\Phi_+$ and $\Phi_-$ are appropriate linear combinations of $\Phi_1$ and $\Phi_2$.
 The envelope amplitude-vector, $\alpha_\star(\zeta)=(\alpha_{\star,+}(\zeta),\alpha_{\star,-}(\zeta))^T$,  is a zero-energy eigenstate, $\mathcal{D}\alpha_\star=0$,  of the 1D Dirac operator: 
 \begin{equation}
 \mathcal{D} \equiv -i|\lambda_\sharp||\ktilde_2|\sigma_3\D_\zeta + \vartheta_\sharp\kappa(\zeta)\sigma_1\ .  \label{dirac_op}
 \end{equation}
Here, $\sigma_1$ and $\sigma_3$ are standard Pauli matrices.

\item Topological protection: The Dirac operator $\mathcal{D}$ has a spatially localized zero-energy eigenstate for any $\kappa(\zeta)$ having asymptotic limits of opposite sign at $\pm\infty$. Therefore, the zero-energy eigenstate, which seeds the bifurcation, persists for \underline{localized} perturbations of $\kappa(\zeta)$. In this sense, the bifurcating branch of edge states is topologically protected against a class of local (not necessarily small) perturbations of the edge.

\item Edge states, $\Psi(\bx;\kpar)\in H^2_{\kpar}$, exist for all parallel quasimomenta $\kpar$ in a neighborhood of $\kparv$, and by symmetry for all $\kpar$ in a neighborhood of $\kpar=-\bK\cdot\vtilde_1=\bK'\cdot\vtilde_1$. It follows that by taking a continuous superposition of the time-harmonic states, $\Psi(\bx,\kpar)e^{-iE(\kpar)t}$, one obtains wave-packets that remain localized about (and dispersing along) the $\vtilde_1-$ edge for all time.
\end{enumerate}
\end{theorem}
We next apply Theorem \ref{general-conditions} to the study of  edge states which are localized transverse to a zigzag edge.

\subsection{Topologically protected zigzag edge states}
\label{zigzag-summary}

The zigzag edge corresponds to the choice: $\vtilde_1=\bv_1$, $\vtilde_2=\bv_2$, and $\ktilde_1=\bk_1$, $\ktilde_2=\bk_2$. In Section 8 of  \cite{FLW-JAMS:15} we apply our general Theorem \ref{general-conditions} to study the zigzag edge eigenvalue problem
$H^{(\eps,\delta)}\Psi = E\Psi$, $\Psi\in H^2_{\kpar}(\Sigma)$, where $\Sigma=\R^2/\Z\bv_1$, 
 for the Hamiltonian \eref{schro-domain}
with:
\begin{equation}
0<|\delta|\lesssim \eps^2 \ll1 .
\label{small-eps-delta}\end{equation}
There are two cases, which are delineated by the sign of the distinguished Fourier coefficient of the unperturbed (bulk) honeycomb potential:
\[\textrm{(1) $\eps V_{1,1}>0$\ \ \  and\ \ \  (2) $\eps V_{1,1}<0$,}\]
where $V_{1,1}$, defined in \eref{V11eq0-intro}, is assumed to be non-zero. For appropriate choices of $V(\bx)$, it is possible to tune been cases (1) and (2) by varying of the lattice scale parameter; see  Appendix A of \cite{FLW-JAMS:15}.

\begin{remark}\label{computer_sims}
The computer simulations discussed in this and subsequent sections were done for the Hamiltonian $H^{(\eps,\delta)}$ with
$\kappa(\zeta)=\ \tanh(\zeta)$ and 
\begin{equation}
\eqalign{
V(\bx) = \sum_{\bk_j\in\{\bk_1,\bk_2,\bk_1+\bk_2\}}\cos(\bk_j \cdot\bx),\\
W(\bx) = \sum_{\bk_j\in\{\bk_1,\bk_2,\bk_1+\bk_2\}} \sin(\bk_j \cdot\bx).}
\label{VW-numerics}
\end{equation}
Here, $\bk_1$ and $\bk_2$ are displayed in Section \ref{honeycomb_potentials}.
Since $V_{1,1}>0$, $\sgn( \eps V_{1,1} )$ is determined by the sign of $\eps$.
\end{remark}

\nit {\bf Case (1) $\eps V_{1,1}>0$ and  \eref{small-eps-delta}:} In this case, the spectral no-fold condition holds for the zigzag edge (Theorem 8.2 of \cite{FLW-JAMS:15}) and it follows that there exist zigzag edge states (Theorem 8.5 of \cite{FLW-JAMS:15}). In particular,  for all $\eps$ and $\delta$ satisfying \eref{small-eps-delta}, the zigzag edge state eigenvalue problem \eref{EVP} has topologically protected edge states, described in Theorem \ref{general-conditions},  with $L^2_\kpar-$ spectra
of energies:  $\kpar\mapsto E(\kpar)$ ($\kpar$ varying near $\bK\cdot\bv_1=2\pi/3$ and varying near $-\bK\cdot\bv_1=-2\pi/3$) sweeping out a neighborhood of $E=E_\star$. 

\begin{figure}
\centering
\includegraphics[width=\columnwidth]{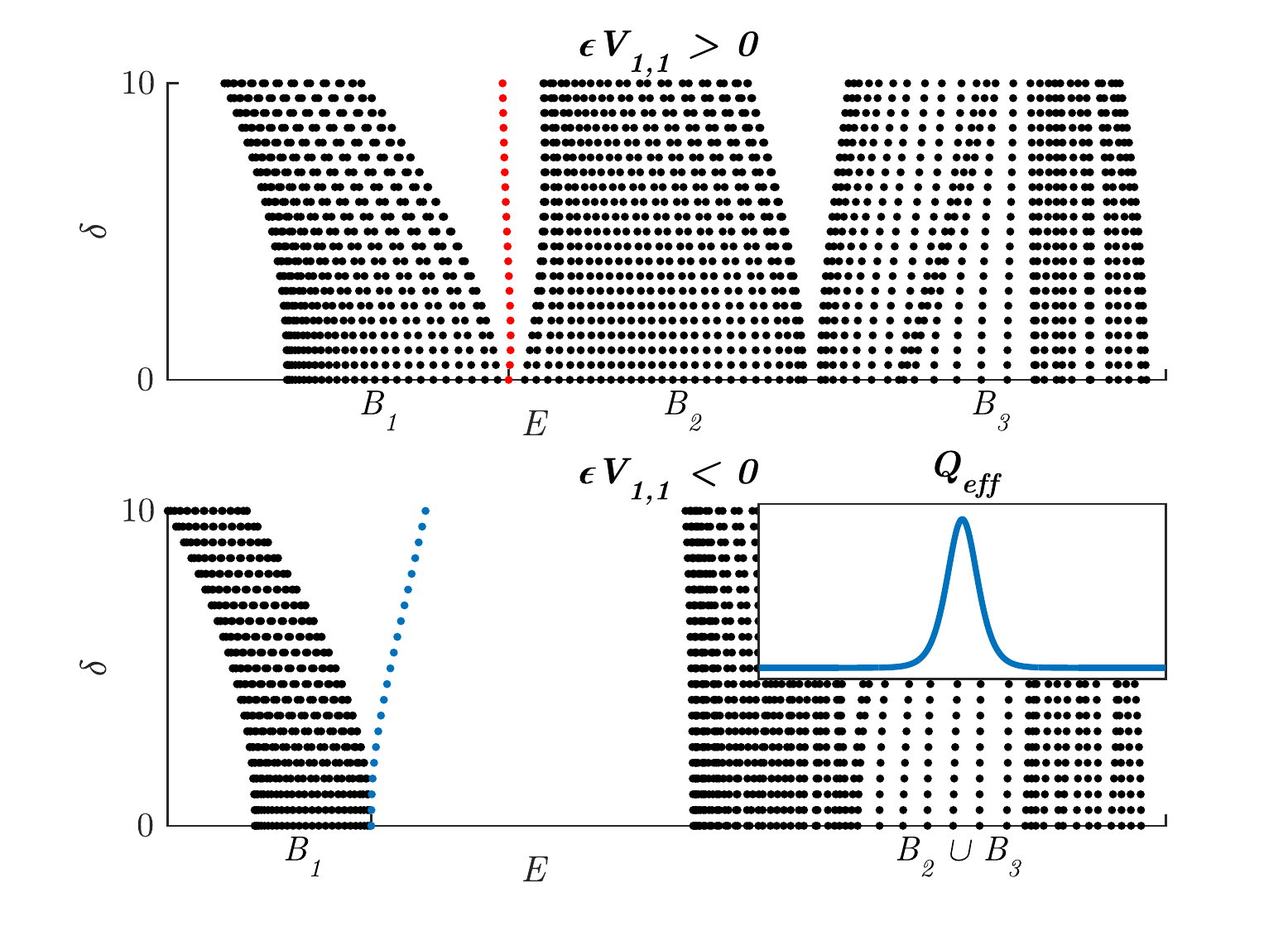}
\caption{Bifurcation of zigzag edge states from spectral band edges of $H^{(\eps,\delta)}$, displayed via a plot of 
$L^2_{\kpar=\bK\cdot\bv_1}-$ spectra vs. $\delta$.  Here, $\bK\cdot\bv_1=\frac{2}{3}\pi$.
 {\bf Top panel:} $\eps V_{1,1}>0$. Topologically protected bifurcation of zigzag edge states (dotted red curve), is seeded by the zero-energy mode of the Dirac operator \eref{dirac_op}. The branch of edge states emanates from the intersection of first and second bands ($B_1$ and $B_2$) at $E=E_\star$ for $\delta=0$.
   {\bf Bottom panel:} $\eps V_{1,1}<0$. Bifurcation of zigzag edge states from endpoint of the first spectral band, $E=\widetilde{E}_\star$. This bifurcation is governed by an effective Schr\"odinger equation \eref{schro_solvability}.
\label{E_delta}
}
\end{figure}

\begin{figure}
\centering
\includegraphics[width=\columnwidth]{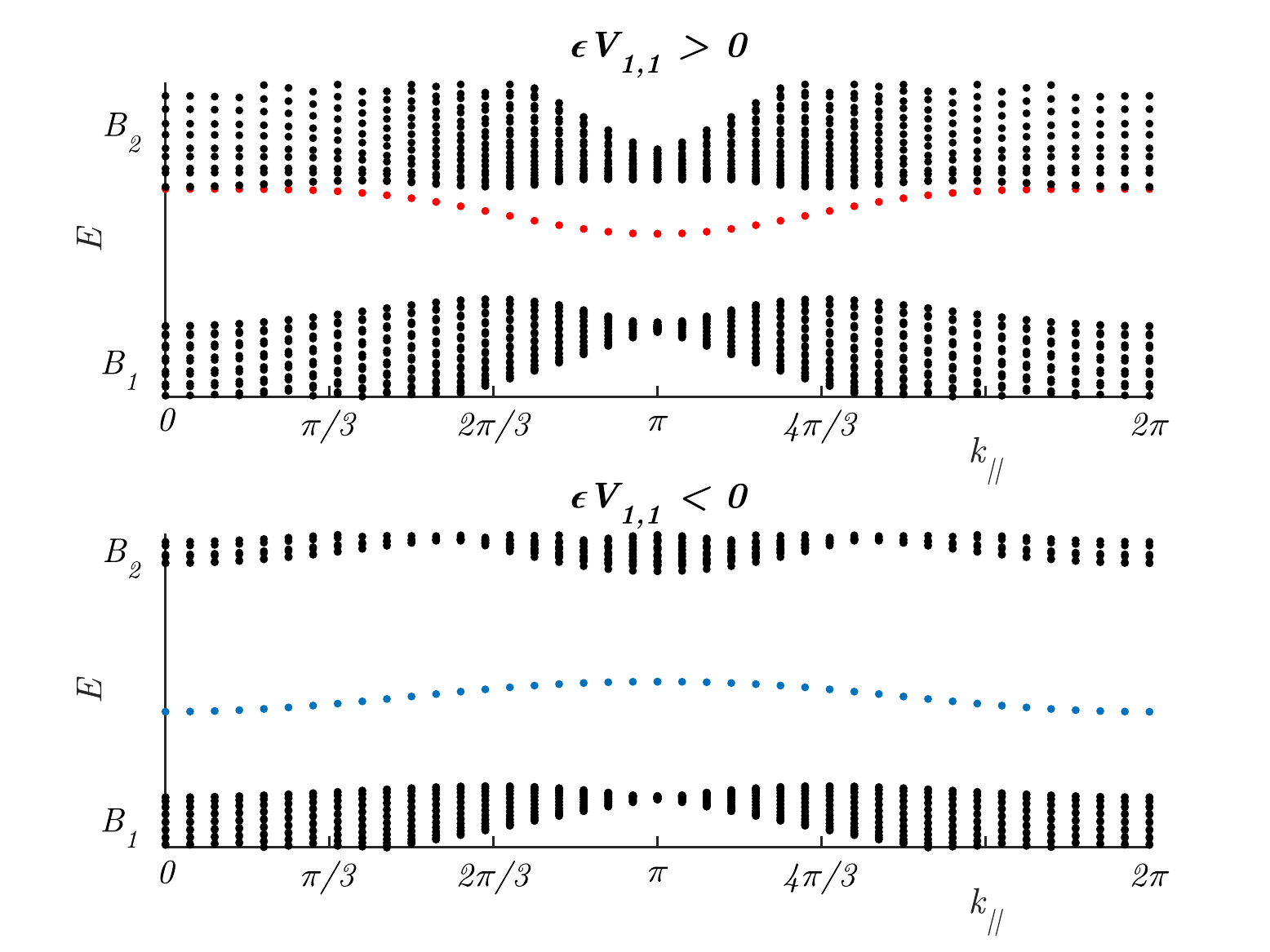}
\caption{
 Energy $L^2_{\kpar}-$ spectrum of $H^{(\eps,\delta)}$ vs. parallel quasimomentum, $\kpar$,   for $\eps V_{1,1}>0$ ({\bf top panel}) and $\eps V_{1,1}<0$ ({\bf bottom panel}).
 \label{fig_zigzag_E_Kpar}}
\end{figure}

Case (1) is illustrated by the top panels of Figures \ref{E_delta} and \ref{fig_zigzag_E_Kpar}.
The top panel of Figure \ref{E_delta} displays, for $\eps=10$,  the $L^2_{\kparpi}-$ spectra (plotted horizontally) of $H^{(\eps,\delta)}$ corresponding to a range of $\delta$ values for the case $\eps V_{1,1}>0$. 
Figure  \ref{fig_zigzag_E_Kpar} displays the $L^2_{\kpar}$ spectra (plotted vertically),
 for fixed $\eps$ and $\delta$,  for a range of  $\kpar$.  

\nit {\bf Case (2) $\eps V_{1,1}<0$ and \eref{small-eps-delta}:} The spectral no-fold condition does not hold  (Theorem 8.4 of \cite{FLW-JAMS:15}). This is clearly illustrated in the top, middle panel of Figure \ref{fig_spectral_nofold}. Therefore, Theorem \ref{general-conditions} does not apply to give the existence of topologically protected edge states. 

On the other hand,  in \cite{FLW-JAMS:15} we give a formal multiple scale perturbation expansion  in powers of $\delta$ (using the procedure of  Section \ref{multiscale}) which solves the edge state eigenvalue problem to any order in $\delta$. We believe, however, that this expansion is not the expansion of a genuine edge state. Rather, we conjecture that it is an approximation of a very long-lived  ``edge quasi-mode''. Such modes would have complex frequency, $E$,  near $E_\star$,  with small imaginary part and would decay on a very long time scale; see  \cite{FLW-JAMS:15}  for further discussion. 

Case (2), $\eps V_{1,1}<0$,  is the point of departure for our discussion of {\it non-protected edge states}, in the following section.

\section{Edge states which are not topologically protected}\label{unprotected}

As noted above, the spectral no-fold condition fails if $\eps V_{1,1}<0$. Theorem \ref{general-conditions} does not yield an edge state for the zigzag edge, although there is evidence of a very long-lived quasi-mode. Note also, by Theorem \ref{generic-dirac-points}, that if $\eps V_{1,1}<0$, Dirac points occur at the intersection of the second and third spectral bands; see the center panel of the top row in Figure \ref{fig_spectral_nofold}.
  
Note further from this same plot that there is an $L^2_{\kpar=\bK\cdot\bv_1}-$ spectral gap between the first and second bands. Numerical computations, displayed in the bottom row of plots in Figures \ref{E_delta} and \ref{fig_zigzag_E_Kpar} show zigzag edge states, for the full range of parallel quasimomenta, $0\le\kpar\le2\pi$, bifurcating from the $L^2_\kpar-$ band edge into the first finite spectral gap. A bifurcation with a similar nature is discussed in \cite{Thorp-etal:15}; see also  \cite{plotnik2013observation}.
    
In the following subsection we present a formal multiple scale perturbation derivation of these bifurcation curves.  In contrast to the  edge state bifurcations obtained via Theorem \ref{general-conditions}, these bifurcations are not topologically protected;  they  may be destroyed by a  localized perturbation of the edge. 
The formal multiple-scale bifurcation analysis presented in Section \ref{multiscale} can be made rigorous along the lines of \cite{ilan2010band,hoefer-weinstein:11,dvw-sima:15,DVW-CMS:15}; see also \cite{Borisov-Gadylshin:08}.

\subsection{Multiple scale perturbation analysis}\label{multiscale}
%
%
For $\bk\in\brill_h$, let $(\widetilde{E}(\bk),\widetilde{\Phi}(\bx;\bk))$ denote the eigenpair associated with a lowest $L^2_{\kparv}-$ spectral band of $H^{(\eps,0)}$. Note that the quasimomentum slice of the band structure, dual to the zigzag edge, is displayed in the middle plot of the top row of Figure \ref{fig_spectral_nofold}. 
The maximum energy of this band, or equivalently the rightmost edge of band $B_1$ in the bottom panel of Figure \ref{E_delta}, is attained for the eigenpair with energy $\widetilde{E}_\star \equiv \widetilde{E}(\bK)$ and Floquet-Bloch mode  $\widetilde{\Phi}_\star(\bx)\equiv \widetilde{\Phi}(\bx;\bK)$  at the quasimomentum $\bk=\bK$. 
More generally, we may consider any vertex quasimomenta, $\bK_\star$, such that the $L^2_{\bK_\star}-$ nullspace of $H^{(\eps,\delta)}-\widetilde{E}_\star I$ has dimension one, which is the case in the numerical examples we have considered.

We keep the discussion general and consider a general $\vtilde_1-$ edge (recall the setup at the beginning of Section \ref{edge_model}) and seek a solution  of the $\vtilde_1-$ edge state eigenvalue problem \eref{edge-evp}-\eref{edge-bc2} (see also \eref{EVP}) of the multi-scale form: $\Psi=\Psi(\bx,\zeta)$ with $\zeta=\delta\ktilde_2\cdot\bx$. 
 In terms of these variables:
\begin{equation}
\label{multi-Schroedinger}
\eqalign{
&\left[-\left(\nabla_\bx+\delta\ktilde_2\partial_\zeta\right)^2+ \eps V(\bx) + \delta\kappa(\zeta)W(\bx)\right] \Psi(\bx;\zeta) \\ 
&\qquad\qquad = E\  \Psi(\bx;\zeta)\\
&\Psi(\bx+\vtilde_1,\zeta)=e^{i\bK_\star\cdot\vtilde_1}\Psi(\bx,\zeta),\\
&\zeta\mapsto\Psi(\bx,\zeta)\in L^2(\R_\zeta).}
\end{equation}
We seek a solution to \eref{multi-Schroedinger} via a multiple scale expansion in $\delta$, assumed to be small:
\begin{equation}
\label{formal-psi}
\eqalign{
E^{\delta}&=E^{(0)}+\delta E^{(1)}+\delta^2E^{(2)}+\ldots, \\
\Psi^{\delta}&= \Psi^{(0)}(\bx,\zeta)+\delta\Psi^{(1)}(\bx,\zeta)+\delta^2\Psi^{(2)}(\bx,\zeta)+\ldots .}
\end{equation}
The boundary conditions \eref{edge-bc1} and \eref{edge-bc2} are imposed by requiring, for $i\ge0$:
\begin{eqnarray*} 
\Psi^{(i)}(\bx+\vtilde_1,\cdot)=e^{i\bK_\star\cdot\vtilde_1}\Psi^{(i)}(\bx,\cdot),\\
\zeta\to\Psi^{(i)}(\bx,\zeta)\in L^2(\R_\zeta).
\end{eqnarray*}
Substituting \eref{formal-psi} into \eref{multi-Schroedinger} and equating terms of equal order in $\delta^i,\ i\ge0$, yields a hierarchy of non-homogeneous boundary value problems, governing $\Psi^{(i)}(\bx,\zeta)$. Each equation in this hierarchy is viewed as a PDE with respect to $\bx$, with pseudo-periodic boundary conditions. The solvability conditions for this hierarchy determine the $\zeta-$ dependence of $\Psi^\delta(\bx,\zeta)$.

At order $\delta^0$ we have that $(E^{(0)},\Psi^{(0)})$ satisfy
\begin{equation}
 \label{perturbed_schro_delta0}
 \eqalign{
 \left(-\Delta_{\bx} +\eps V(\bx)-E^{(0)}\right)\Psi^{(0)} = 0, \\
 \Psi^{(0)}(\bx+\bv,\zeta) = e^{i\bK_\star\cdot\bv}\Psi^{(0)}(\bx,\zeta),
  \quad \textrm{for all}\ \bv\in \Lambda_h .}
\end{equation} 
Equation \eref{perturbed_schro_delta0} has solution
\begin{equation}
E^{(0)}=\widetilde{E}_\star,\quad \Psi^{(0)}(\bx,\zeta)=A_0(\zeta)\widetilde{\Phi}_\star(\bx) ,
\label{psi0-soln}
\end{equation}
 where, $\widetilde{\Phi}_\star$ (normalized in $L^2(\Omega)$) spans the nullspace of $H^{(\eps,0)}-\widetilde{E}_{\star}I$.

Proceeding to order $\delta^1$ we find that $(\Psi^{(1)},E^{(1)})$ satisfies
\begin{equation}
 \label{psi1-eqn}
 \eqalign{
 &\left(-\Delta_{\bx}+\eps V(\bx)-E_{\star}\right)\Psi^{(1)}(\bx,\zeta)=G^{(1)}(\bx,\zeta)\\
 &\Psi^{(1)}(\bx+\bv,\zeta) = e^{i\bK_\star\cdot\bv}\Psi^{(1)}(\bx,\zeta), \quad \textrm{for all}\ \bv\in \Lambda_h,}
\end{equation}
 where
 \begin{equation}
 \label{G1def}
 \eqalign{
 G^{(1)}(\bx,\zeta)&=G^{(1)}(\bx,\zeta;\Psi^{(0)},E^{(1)}) \\
 &\equiv 2\partial_\zeta A_0(\zeta)\ \ktilde_2 \cdot\nabla_\bx\widetilde{\Phi}_\star(\bx) \\
 &\quad + \left(-\kappa(\zeta)W(\bx)+E^{(1)}\right)A_0(\zeta)\widetilde{\Phi}_\star(\bx) .}
\end{equation}
A necessary condition for the solvability of \eref{psi1-eqn} is that the inhomogeneous term on the right hand side be  $L^2_{\bK_\star}(\Omega;d\bx)-$ orthogonal to 
$\widetilde{\Phi}_\star(\bx)$. 
By the symmetries $\widetilde{\Phi}_\star(\bx) = \overline{\widetilde{\Phi}_\star}(-\bx)$ and $W(\bx)=-W(-\bx)$,
\begin{eqnarray}
&\inner{\widetilde{\Phi}_\star,W \widetilde{\Phi}_\star}_{L^2_{\bK_\star}(\Omega)}= 0, \ \  
 \inner{\widetilde{\Phi}_\star,\nabla\widetilde{\Phi}_\star}_{L^2_{\bK_\star}(\Omega)}=0. \label{inner_zero} 
\end{eqnarray}
Therefore, with $E^{(1)}\equiv0$,  the right hand side of \eref{psi1-eqn} lies in the range of $H^{(\eps,0)}-\widetilde{E}_\star I: H^2_{\bK_\star}\to L^2_{\bK_\star}$, and we  obtain
\begin{equation}
 \Psi^{(1)}(\bx,\zeta) = \left(R(E_{\star})G^{(1)}\right)(\bx,\zeta;0) , \label{psi1p-def}
\end{equation} 
where 
$R(\widetilde{E}_\star) = (H^{(\eps,0)}-\widetilde{E}_\star I)^{-1}:P_\perp L^2(\Omega)\to H^2(\Omega)$.
Here, $P_\perp$ is the $L^2-$  projection onto ${\rm span}\{\widetilde{\Phi}_\star\}^\perp$.

At order $\delta^2$, we have
\begin{eqnarray}
 \label{psi2-eqn}
 \eqalign{
& \left(-\Delta_\bx +\eps V(\bx)-\widetilde{E}_\star\right)\Psi^{(2)}(\bx,\zeta)
 = G^{(2)}(\bx,\zeta) \\
&\Psi^{(2)}(\bx+\bv,\zeta) = e^{i\bK_\star\cdot\bv}\Psi^{(2)}(\bx,\zeta), \quad \textrm{for all}\ \bv\in \Lambda_h ,}
\end{eqnarray}
where
\begin{eqnarray*}
 G^{(2)}(\bx,\zeta) &= G^{(2)}(\bx,\zeta;\Psi^{(0)},\Psi^{(1)},E^{(2)}) \nn \\
 &\equiv\left(2\nabla_{\bx}\cdot \ktilde_2\partial_\zeta-\kappa(\zeta)W(\bx)\right)\psi_p^{(1)} \\ 
&\quad + |\ktilde_2|^2\partial_\zeta^2 A_0(\zeta)\widetilde{\Phi}_\star(\bx) 
 + E^{(2)}A_0(\zeta)\widetilde{\Phi}_\star(\bx). \nn
\end{eqnarray*}
Equation \eref{psi2-eqn} has a solution if and only if the right hand side is $L^2_{\bK_\star}$-orthogonal to $\widetilde{\Phi}_\star$.
This solvability condition (all inner products over $L^2_{\bK_\star}(\Omega_\bx)$) is:
\begin{eqnarray}
&|\ktilde_2|^2 \partial_\zeta^2 A_0(\zeta) + \inner{\widetilde{\Phi}_\star(\bx),2\nabla_\bx\cdot\ktilde_2\partial_\zeta \Psi^{(1)}(\bx,\zeta)} \label{delta2-solvability} \\
&\qquad -\inner{\widetilde{\Phi}_\star(\bx), \kappa(\zeta) W \Psi^{(1)}(\bx,\zeta)} + E^{(2)} A_0(\bx) = 0.\nn
\end{eqnarray}

We  simplify \eref{delta2-solvability} using the expression for $\Psi^{(1)}$ in \eref{psi1p-def}; see also \eref{G1def}.
Importantly, note that the action of the resolvent $R(\widetilde{E}_\star)$ on the individual terms in \eref{G1def} is well defined due to \eref{inner_zero}.
After some manipulation, the first inner product in \eref{delta2-solvability} yields
\begin{eqnarray}
&\inner{\widetilde{\Phi}_\star(\bx),2\nabla_\bx\cdot\ktilde_2\partial_\zeta \Psi^{(1)}(\bx,\zeta)} \nn \\ 
& = -4 \sum_{1\leq i,j \leq 2} 
\inner{\partial_{x_i}\widetilde{\Phi}_\star(\bx),  R(\widetilde{E}_\star)(\partial_{x_j}\widetilde{\Phi}_\star)} \ktilde_2^{(i)} \ktilde_2^{(j)} \partial_{\zeta}^2A_0(\zeta)  \nn \\
&\qquad  +2 \inner{\nabla_\bx\widetilde{\Phi}_\star, R(\widetilde{E}_\star)(W\widetilde{\Phi}_\star)} \cdot \ktilde_2\partial_\zeta (\kappa(\zeta) A_0(\zeta)) , \label{first_inner}
\end{eqnarray}
while the second inner product in \eref{delta2-solvability} yields
\begin{eqnarray}
&-\inner{\widetilde{\Phi}_\star(\bx), \kappa(\zeta) W(\bx) \Psi^{(1)}(\bx,\zeta)} \nn \\
& = -2\inner{\widetilde{\Phi}_\star, W R(\widetilde{E}_\star) (\nabla_\bx \widetilde{\Phi}_\star)} \cdot \ktilde_2 \kappa(\zeta) \partial_\zeta A_0(\zeta)\nn \\ 
&\qquad + \inner{\widetilde{\Phi}_\star, W R(\widetilde{E}_\star) (W \widetilde{\Phi}_\star)} \kappa^2(\zeta) A_0(\zeta) .
\label{second_inner}
\end{eqnarray}
It can be checked that 
\begin{eqnarray}
 \inner{\widetilde{\Phi}_\star, W R(\widetilde{E}_\star) (W \widetilde{\Phi}_\star)}_{L^2_{\bK_\star}(\Omega)}  > 0 \label{positive} \quad \textrm{and that} \\
 \inner{\widetilde{\Phi}_\star, W R(\widetilde{E}_\star) (\nabla_\bx \widetilde{\Phi}_\star)}_{L^2_{\bK_\star}(\Omega_\bx)}\ \textrm{is real.} \label{realness}
\end{eqnarray}
Relation \eref{positive} follows since $\widetilde{E}_\star$ is the lowest eigenvalue of $H^{(\eps,0)}(\bK_\star)$.
Moreover we have that
\begin{eqnarray}
 &\delta_{ij} |\ktilde_2|^2 -
 4 \sum_{1\leq i,j \leq 2}  \inner{\partial_{x_i}\widetilde{\Phi}_\star(\bx),  R(\widetilde{E}_\star)(\partial_{x_j}\widetilde{\Phi}_\star)} \ktilde_2^{(i)} \ktilde_2^{(j)}  \nn\\
 &\qquad =  \frac{1}{2} [D^2\widetilde{E}_\star(\bK_\star)]_{ij} \ktilde_2^{(i)} \ktilde_2^{(j)} \equiv \frac{1}{2 m_{\rm eff}} , \label{curvature}
\end{eqnarray}
where $D^2\widetilde{E}_\star(\bK_\star)$ is the Hessian matrix of $\widetilde{E}(\bk)$ evaluated at $\bk=\bK_\star$ along the $\ktilde_2-$ quasimomentum direction, and $m_{\rm eff}$ is the effective mass \cite{kittel:96}; see also Theorem 3.1 and equation (3.6) of \cite{ilan2010band}.

Substituting \eref{first_inner} and \eref{second_inner} into equation \eref{delta2-solvability}, and using relations \eref{realness} and \eref{curvature}, we find that the $\mathcal{O}(\delta^2)$ solvability condition reduces to the eigenvalue problem for an effective Schr\"odinger operator, $H_{\rm eff}$:
\begin{equation}
\label{schro_solvability}
\eqalign{
H_{\rm eff} A_0(\zeta) =\mu_{\rm eff} A_0(\zeta), \quad A_0(\zeta)\in L^2(\R_\zeta), \quad \textrm{where} \\
H_{\rm eff} \equiv -\frac{1}{2 m_{\rm eff}} \partial_\zeta^2 + Q_{\rm eff}(\zeta; \kappa),}
\end{equation}
with eigenvalue parameter $\mu_{\rm eff} = E^{(2)} + b\ \kappa_\infty^2$  and  effective potential, $Q_{\rm eff}$,  given by:
\begin{eqnarray}
 &Q_{\rm eff}(\zeta; \kappa) \nn\\
 \ &=  -2\inner{ \ktilde_2\cdot \nabla_\bx \widetilde{\Phi}_\star, R(\widetilde{E}_\star) (W \widetilde{\Phi}_\star)}_{L^2_{\bK_\star}(\Omega)}
  \kappa'(\zeta) \nn \\
 &\quad + \inner{W\widetilde{\Phi}_\star, R(\widetilde{E}_\star) (W \widetilde{\Phi}_\star)}_{L^2_{\bK_\star}(\Omega)} \left(\kappa^2_\infty-\kappa^2(\zeta) \right)  \nn \\
 &\equiv a\ \kappa'(\zeta) + b \left(\kappa^2_\infty-\kappa^2(\zeta) \right). \label{Qeff}
\end{eqnarray}
The constants $a$ and $b$ depend on $\widetilde{\Phi}_\star, V$ and $W$ and, by \eref{positive}, $b>0$. 
Since $\kappa(\zeta)$ satisfies $\kappa(\zeta)\to\pm\kappa_\infty$ as $\zeta\to\pm\infty$, both $\kappa'(\zeta)$ and $\kappa^2_\infty-\kappa^2(\zeta)$ decay to zero as $\zeta\to\pm\infty$. Therefore, $Q_{\rm eff}(\zeta; \kappa)$ is decaying at infinity.

If the eigenvalue problem \eref{schro_solvability} has a non-trivial solution, then  the above formal expansion yields a solution of the $\vtilde_1-$ edge state eigenvalue problem to arbitrary finite order in $\delta$. This formal expansion yields approximations to arbitrary finite order of genuine solutions of the $\vtilde_1-$ edge state eigenvalue problem, $\delta\mapsto (\widetilde{\Psi}^\delta(\bx),\widetilde{E}^\delta)$.  

In summary, the non-trivial eigensolutions of the effective Schr\"odinger operator, $H_{\rm eff}$, seed or induce the bifurcation of non-trivial edge states in a manner analogous to the zero-energy eigenmodes of a Dirac operator inducing the bifurcation of protected edge states from Dirac points. 
 In the next section we show that such ``effective Schr\"odinger bifurcations'', also known as ``Tamm states'',  are not topologically protected; these states do not persist in the presence of arbitrary localized perturbations of the domain wall.

\begin{figure}
\centering
\includegraphics[width=\columnwidth]{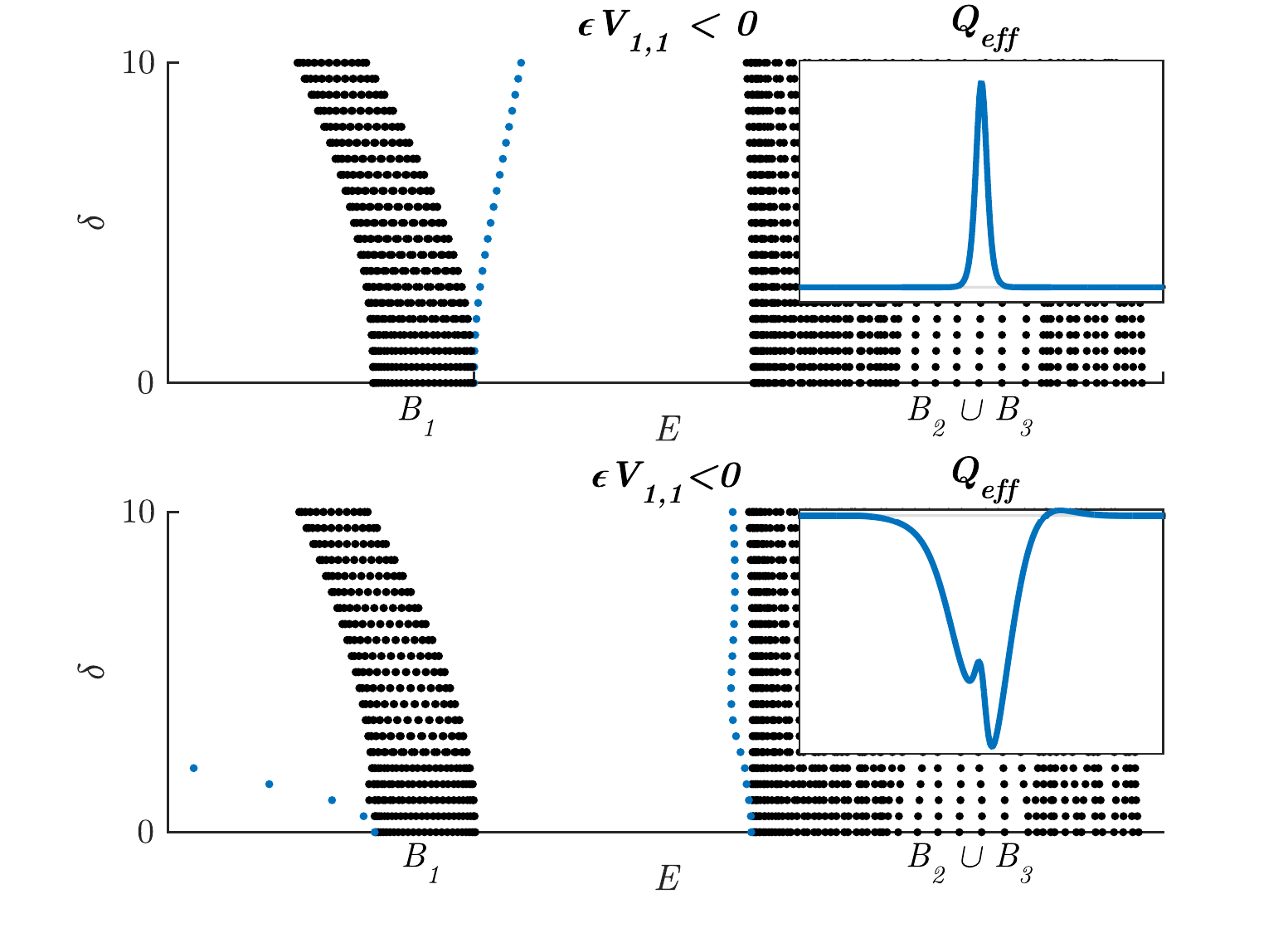}
\caption{ Bifurcation curves of non-protected edge states from spectral band edges, encoded in 
$L^2_{\kparpi}-$ spectra vs. $\delta$ for the Hamiltonian $H^{(-10,\delta)}$ for $\eps V_{1,1}<0$. 
{\bf Top panel}: $\kappa(\zeta)=\tanh(\zeta)$. 
{\bf Bottom panel}: $\kappa_{\natural}(\zeta)=\kappa(\zeta)+F(\zeta)= \tanh(\zeta) + 10 \exp(-\zeta^2/50)$.
The branch of edge states bifurcating from the top edge of the lowest band, $B_1$, at $E=\widetilde{E}_\star$ for $\delta=0$ (blue dotted curve in top panel) is destroyed by the perturbation $F(\zeta)$.
Insets show the effective potential $Q_{\rm eff}$ in each case.}
 \label{fig_E_delta_perturbed}
\end{figure}

\subsection{Edge states seeded by localized eigenstates of the effective Schr\"odinger operator,  $H_{\rm eff}$,  are \underline{not} topologically protected}
\label{not-protected}

We now explore the robustness of the edge state bifurcating from the trivial state $E=\widetilde{E}_\star$ for $\delta=0$, constructed in Section \ref{multiscale}.
We focus on the case of the zigzag edge, corresponding to the choice $\vtilde_1=\bv_1$, $\vtilde_2=\bv_2$ and $\ktilde_1=\bk_1$, $\ktilde_2=\bk_2$. We also fix $V$, $W$ and $\kappa(\zeta)=\tanh(\zeta)$ as in \eref{VW-numerics}, with $\eps=-10$ so that $\eps V_{1,1}<0$. From numerics we have $m_{\rm eff}^{-1}<0$ (this is also clear from the top, center panel of Figure \ref{fig_spectral_nofold} and Figure \ref{fig_E_delta_perturbed}). Moreover, $a>0$ and, as argued above, $b>0$.
Therefore, $\kappa'(\zeta)=\sech^2(\zeta)$ and $\kappa_\infty^2-\kappa^2(\zeta)=\sech^2(\zeta)$ are non-negative and it  follows that $Q_{\rm eff}(\zeta;\kappa)$ \eref{Qeff}, plotted in the inset of Figure \ref{E_delta} (and the top inset of Figure \ref{fig_E_delta_perturbed}), is a potential barrier. Therefore, since  $m_{\rm eff}^{-1}<0$, $H_{\rm eff}$ has a positive energy eigenstate. This eigenstate induces a bifurcation curve of edge states, emanating from the upper edge of $B_1$, into the spectral gap to its right; see the top panel of Figure \ref{fig_E_delta_perturbed}.

We next construct a domain wall function, $\kappa_\natural(\zeta)$,  for which the effective potential, $Q_{\rm eff}(\zeta;\kappa_\natural)$, is such that $H_{\rm eff}=-(2 m_{\rm eff})^{-1}\partial_\zeta^2 + Q_{\rm eff}(\zeta; \kappa_\natural)$, with $m_{\rm eff}^{-1}<0$ does not have a positive energy eigenstate.  Therefore,  
no bifurcation from the upper edge of $B_1$ into the gap above is expected.  Explicitly, let  $\kappa_\natural(\zeta) = \kappa(\zeta) + F(\zeta)$, where $F(\zeta)=10 \exp(-\zeta^2/50)$. The bottom panel in Figure \ref{fig_E_delta_perturbed} shows the $L^2_{\kparpi}-$ spectra of $H^{(-10,\delta)}$ for the perturbed domain wall function $\kappa_\natural(\zeta)$. The bifurcating branch emanating from $E=\widetilde{E}_\star$ (the right edge of the first spectral band, $B_1$) has been destroyed by the perturbation $F(\zeta)$. $Q_{\rm eff}(\zeta;\kappa_\natural)$ is plotted in the bottom inset of Figure \ref{fig_E_delta_perturbed}.

The one-parameter family of effective potentials, $Q_{\rm eff}(\zeta; (1-\theta)\kappa+\theta\kappa_{\natural})$, $0\leq\theta\leq1$, provides a smooth homotopy from a Hamiltonian for which a branch of edge states bifurcates from the upper edge of the first spectral band ($H^{(\eps,\delta)} $ with domain wall $\kappa$) to a Hamiltonian for which the branch of edge states does not exist ($H^{(\eps,\delta)} $ with domain wall $\kappa_\natural$). Therefore, this type of bifurcation is \emph{not} topologically protected.
%
%
This contrast between topologically protected and non-protected states is discussed  in an analogous setting of 1D photonic structures in \cite{Thorp-etal:15}.

\begin{figure}
\centering
\includegraphics[width=\columnwidth]{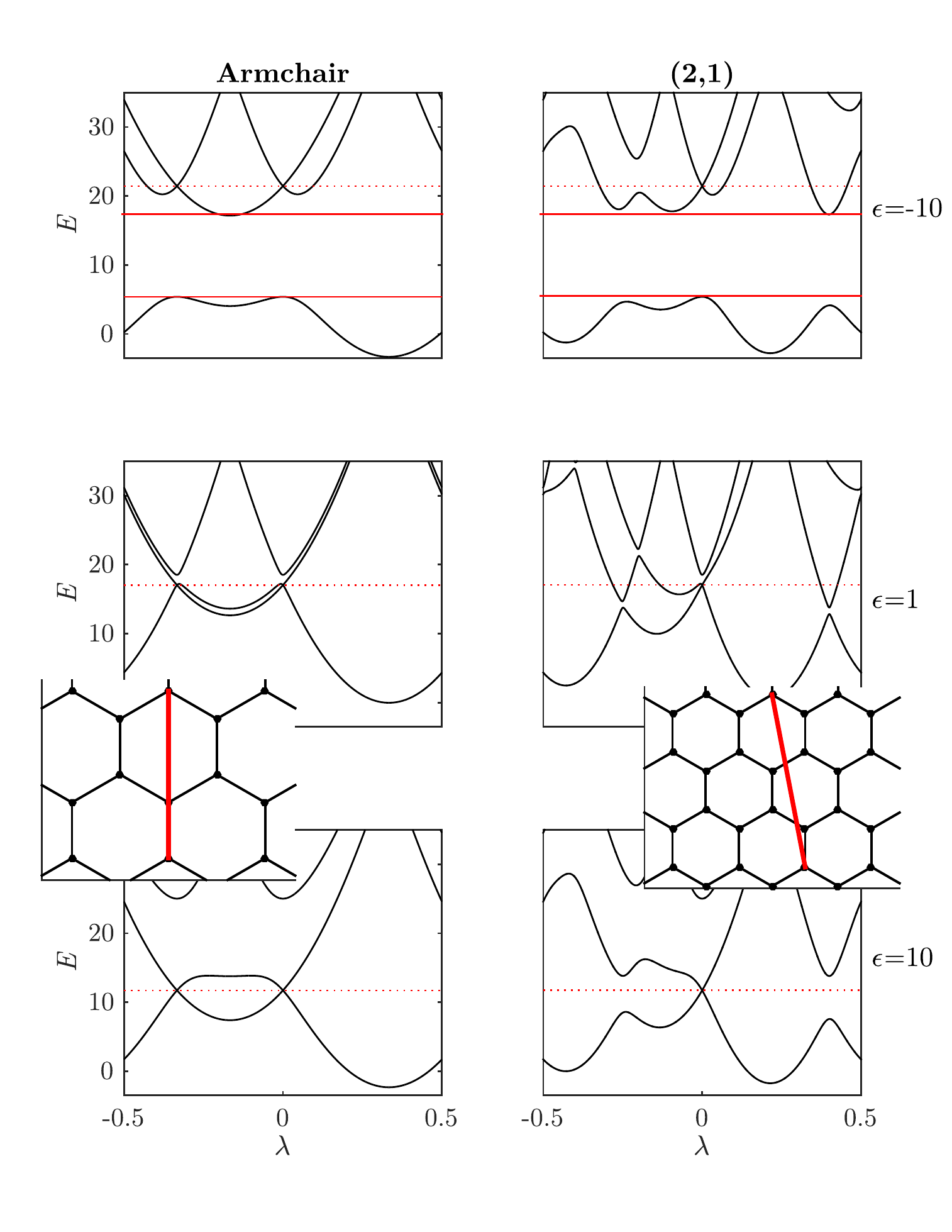}
\caption{
Band dispersion slices of $H^{(\eps,0)}$ along the quasimomentum segments:  $\bK+\lambda\ktilde_2,\ |\lambda|\le1/2$, for $\ktilde_2=-\bk_1+\bk_2$ (armchair) and $\ktilde_2=-\bk_1+2\bk_2$ ($(2,1)$), for various fixed $\eps$.
Energy levels $E=E_\star$ (for $\eps>0$) and $E=\widetilde{E}_\star$ (for $\eps<0$) are indicated with dotted lines.
{\bf First row}: $\eps=-10$. A spectral gap exists between the first and second bands.
{\bf Middle row}:  $\eps=1$. No spectral gaps.
{\bf Bottom row}: $\eps=+10$. Spectral no-fold condition holds along both the armchair edge and the $(2,1)-$ edge.
Insets indicate armchair and $(2,1)-$ quasimomentum segments (1D Brillouin zones) parametrized by $\lambda$, for $0\leq\lambda\leq1$.
\label{fig_armchair_21_nofold}}
\end{figure}

\begin{figure}
\centering
\includegraphics[width=\columnwidth]{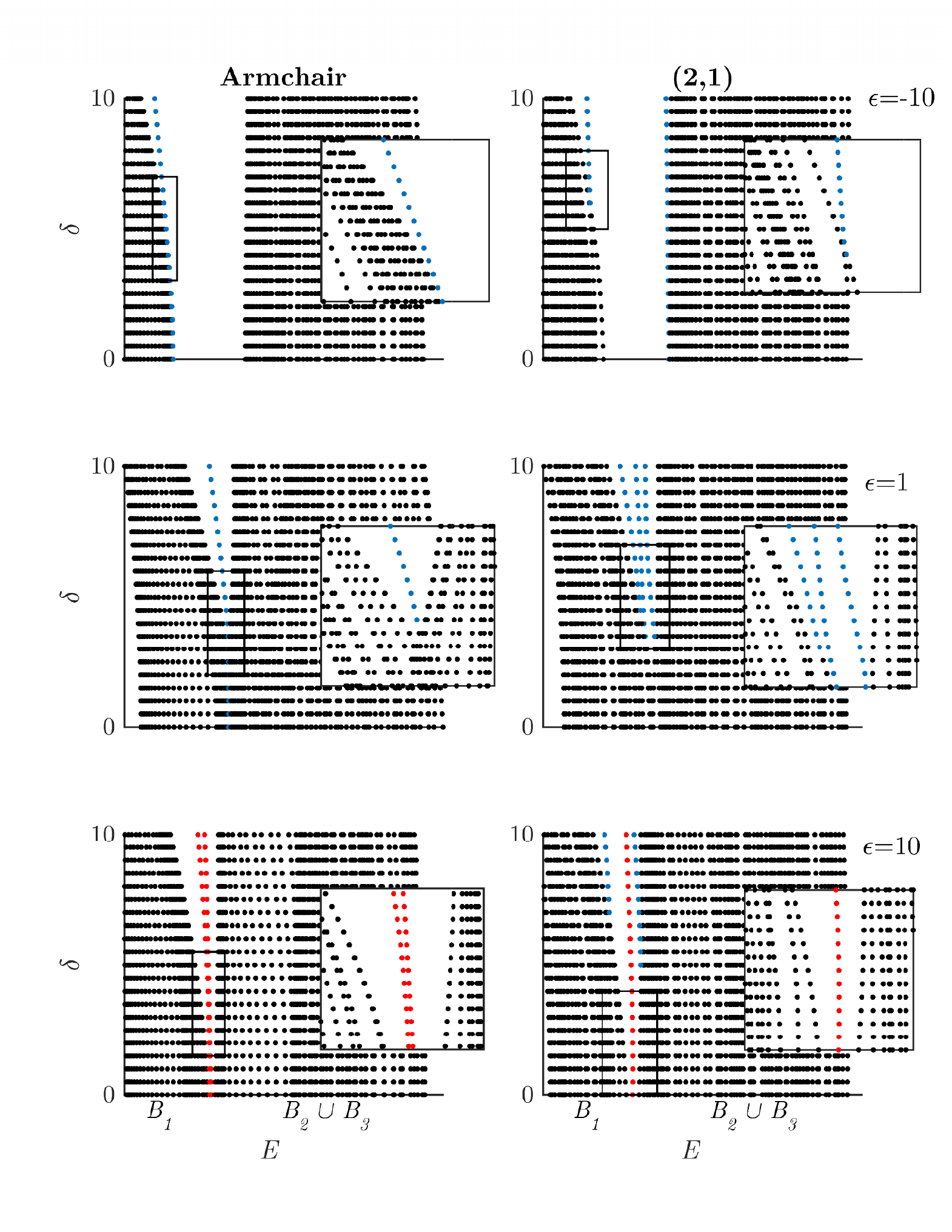}
\caption{ Bifurcation curves of edge states (colored curves), for armchair and ``$(2,1)-$'' edges, at different bulk structure contrasts, encoded as $L^2_{\kparv}-$ spectra of $H^{(\eps,\delta)}$ for various fixed $\eps$. Dark areas are continuous spectra.  Insets zoom in on or near bifurcation points. {\bf First row}: $\eps=-10$. Non-protected edge states (dotted blue curves) bifurcate into the spectral gap between the first and second bands for $\delta>0$.
{\bf Middle row}: $\eps=1$. Non-protected edge states (blue) bifurcate into the spectral gap which opens for $\delta>\delta_\star>0$.
{\bf Bottom row}: $\eps=+10$. Topologically protected (red) and non-protected (blue) edge states bifurcate into the spectral gap for $\delta>0$.
\label{fig_armchair_21_edges_E_delta}}
\end{figure}

\section{Numerical investigation of general edges}\label{numerics_general_edges}

We shall refer to an $(a_1,b_1)-$ edge, or equivalently a $\vtilde_1=a_1\bv_1+b_1\bv_2-$ edge.
In this section we numerically study the edge state eigenvalue problem for the armchair edge ($(1,1)-$ edge) and the $(2,1)-$ edge.  Throughout  we fix $V$ and $W$  as in \eref{VW-numerics} and take $\kappa(\zeta)=\tanh(\zeta)$. The three rows of Figure \ref{fig_armchair_21_nofold} display, for $\eps=-10,1,10$, the band structure slices of $H^{(\eps,0)}(\bK+\lambda\ktilde_2),\ |\lambda|\le1/2$, which are dual to the armchair edge: $\ktilde_2=\ktilde_{AC}=-\bk_1+\bk_2$, and the $(2,1)-$ edge: $\ktilde_2=\ktilde_{(2,1)}=-\bk_1+2\bk_2$. 
Figure \ref{fig_armchair_21_edges_E_delta} displays the corresponding $L^2_{\kparv}(\Sigma)-$ spectra, for $\Sigma=\Sigma_{\rm AC}=\R^2/(\bv_1+\bv_2)$ and 
 $\Sigma=\Sigma_{(2,1)}=\R^2/(2\bv_1+\bv_2)$

\subsection{$\eps=-10$; top rows of Figures \ref{fig_armchair_21_nofold} and \ref{fig_armchair_21_edges_E_delta}}
  Dirac points occur at the intersection of the second and third dispersion surfaces.
(For sufficiently \emph{small} $\eps$,  this is implied by Theorem \ref{generic-dirac-points}, since $\eps V_{1,1} <0$.)

\begin{enumerate}
\item {\bf Armchair edge:} 
Top, left panels of Figures \ref{fig_armchair_21_nofold} and \ref{fig_armchair_21_edges_E_delta}. 
The  band structure slice, associated with the armchair edge, has linear crossings at the Dirac points: $\bK$ ($\lambda=0$) and $\bK'=-\bK$ ($\lambda=2/3$ or, equivalently, $\lambda=-1/3$). 
$H^{(\eps,0)}$ does not satisfy the spectral no-fold condition for the armchair slice; introducing the domain wall/edge potential, with $\delta$ nonzero and small, does not open a full $L^2_{\kpar=0}-$ spectral gap.

\begin{enumerate}[(1)]
 \item Although Theorem \ref{general-conditions}  does not yield an edge state bifurcation from Dirac points at the intersection of the second and third dispersion bands, a formal multiple scale perturbation expansion of an edge state can be constructed as in \cite{FLW-JAMS:15} and suggests that there are  edge quasi-modes, associated with the $\bK$ and $\bK'$ points, which decay exponentially with advancing time.
 
 \item For arbitrary small $\delta\ne0$, edge states ($L^2_{\kpar=0}-$ bound states) bifurcate into the $L^2_{\kpar=0}-$ spectral gap between the first and second bands (Figure \ref{fig_armchair_21_edges_E_delta}). The multiple scale expansion of Section \ref{multiscale} yields edge states which are seeded by the effective Schr\"odinger operator $H_{\rm eff}$ \eref{schro_solvability}. These are not topologically protected; see Section \ref{not-protected}. 

\end{enumerate}

\item {\bf $(2,1)-$ edge:} Top, right panels of Figures \ref{fig_armchair_21_nofold} and \ref{fig_armchair_21_edges_E_delta}. The bifurcation scenario is qualitatively similar to the armchair edge case.

\end{enumerate}

\subsection{$\eps=1$; middle rows of Figures \ref{fig_armchair_21_nofold} and \ref{fig_armchair_21_edges_E_delta}}\label{eps-eq1}

Dirac points occur at the intersection of the first and second dispersion surfaces. 
(For sufficiently small $\eps$ this is implied by Theorem \ref{generic-dirac-points} since  $\eps V_{1,1} >0$.)

\begin{enumerate}

\item {\bf Armchair edge:} 
Middle, left panels of Figures \ref{fig_armchair_21_nofold} and \ref{fig_armchair_21_edges_E_delta}. $H^{(\eps,0)}$ does not satisfy the spectral no-fold condition. Hence, for $\delta$ sufficiently small and non-zero, there is no $L^2_{\kpar=0}-$ spectral gap between the first and second spectral bands.

\begin{enumerate}[(1)]
 \item As in the case $\eps=-10$, the formal multiple scale perturbation theory  suggests the existence of edge quasi-modes with complex frequencies, which decay exponentially with advancing time.
 
 \item Numerical simulations show that there is a threshold value,   $\delta_\star>0$, such that for $\delta>\delta_\star$, there is an $L^2_{\kpar=0}-$ spectral gap between the first and second bands. The edge state branch bifurcating  from a band edge into a gap at a positive threshold value of $\delta$ is shown in Figure  \ref{fig_armchair_21_edges_E_delta}. We believe that this branch is induced by an eigenstate of an effective Schr\"odinger eigenvalue problem, captured by an expansion similar to that in Section \ref{multiscale}, and therefore that it is not topologically protected. 
\end{enumerate}

\item {\bf $(2,1)-$ edge:} Middle, right panels of Figures \ref{fig_armchair_21_nofold} and \ref{fig_armchair_21_edges_E_delta}.
The bifurcation scenario is again qualitatively similar to the armchair edge case. Note that multiple band edge (effective Schr\"odinger) bifurcations occur. This scenario arises when the effective Schroedinger operator has multiple discrete eigenvalues; see \cite{hoefer-weinstein:11}.

\end{enumerate}

\subsection{$\eps=10$;   bottom rows of Figures \ref{fig_armchair_21_nofold} and \ref{fig_armchair_21_edges_E_delta}}\label{eps-eq-10}

This set of simulations reveals additional phenomena and motivate ongoing  mathematical study.
We note that as $\eps$ (bulk material contrast) is increased, the dispersion surfaces begin to separate.
In both armchair edge  and $(2,1)-$ edge cases, Dirac points occur at the intersection of the first and second dispersion surfaces. 

\begin{figure}
\centering
\includegraphics[width=\columnwidth]{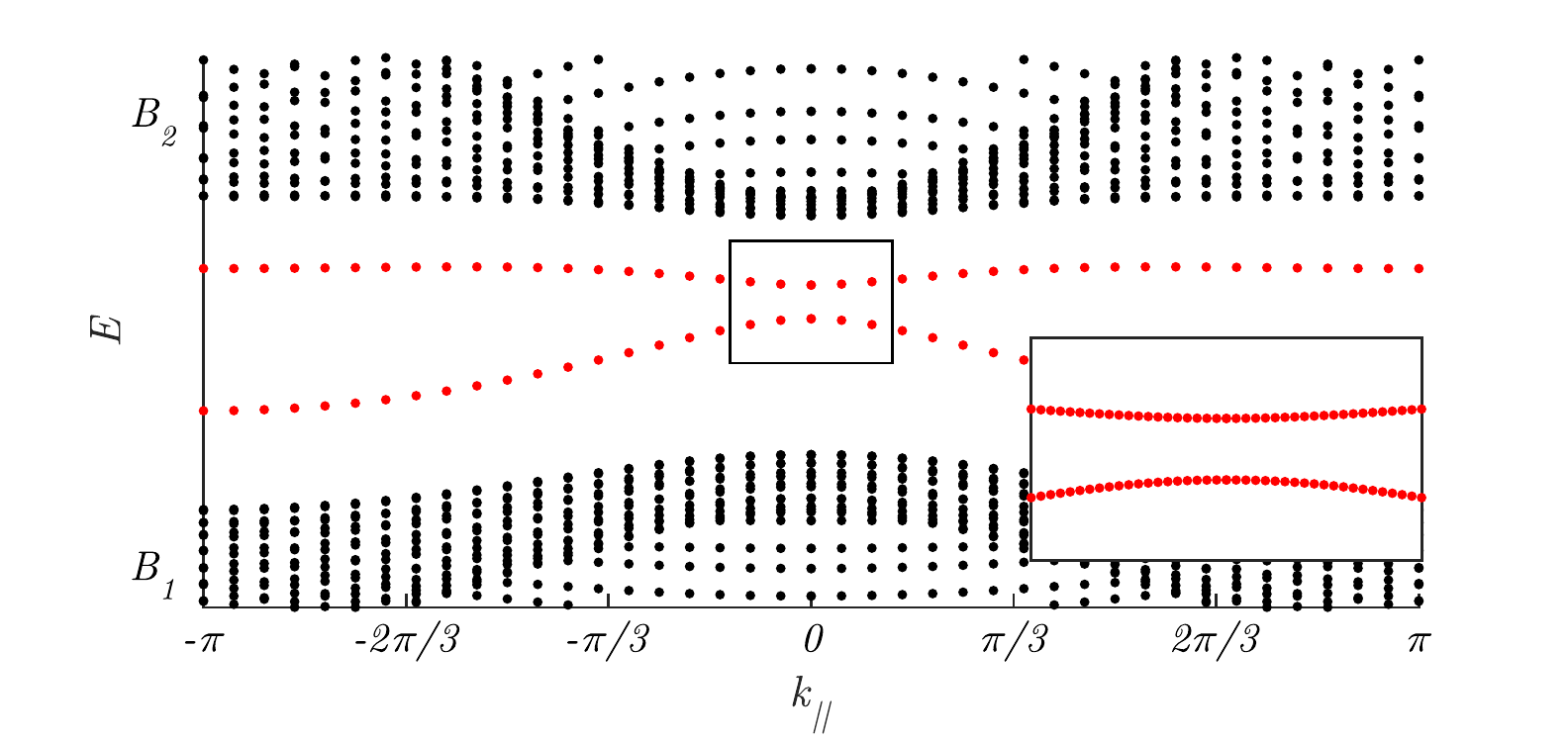}
\caption{
 Bifurcation of two families of armchair edge states, displayed via $L^2_{\kpar}-$ spectrum of $H^{(10,5)}$ vs. $\delta$. For each parallel quasimomentum, $\kpar\in(-\pi,\pi)$, there are two counterpropagating edge states, localized transverse to the armchair edge. 
\label{fig_E_kpar_armchair}}
\end{figure}

\begin{enumerate}

\item {\bf Armchair edge:} 
Bottom, left panels of Figures \ref{fig_armchair_21_nofold} and \ref{fig_armchair_21_edges_E_delta}. 
In Figure \ref{fig_armchair_21_nofold}, we see that along the quasimomentum slice that corresponds to the armchair edge, there are two independent Dirac points, one of $\bK-$ type (in $\bK+\Lambda_h^*$) and one of $\bK'-$ type (in $\bK'+\Lambda_h^*$). The spectral no-fold condition  holds in the more general sense discussed in Section \ref{no-fold}; see bottom, left panel of Figure \ref{fig_armchair_21_nofold}. The results of \cite{FLW-JAMS:15}, as stated, do not apply to give 
the existence of edge states; in \cite{FLW-JAMS:15} it is assumed that the band structure slice, dual to the $\vtilde_1-$ edge intersects at most one Dirac point. However, numerical simulations reveal two families of edge states, associated with $\bK$ and $\bK'$ points, corresponding to counterpropagating armchair edge states; see Figure \ref{fig_E_kpar_armchair}.  
We believe that the analysis of \cite{FLW-JAMS:15} can be extended to give a rigorous construction of these states (article in preparation).  These two branches are seeded by the degenerate zero-energy eigenstate of a $4\times4$ block-diagonal system of Dirac operators.

\item {\bf $(2,1)-$ edge:} Bottom, right panels of Figures \ref{fig_armchair_21_nofold} and \ref{fig_armchair_21_edges_E_delta}. The spectral no-fold condition holds and we expect the general theorem of \cite{FLW-JAMS:15} to apply.
\end{enumerate}

Further numerical studies (not included) of other $(a_1,b_1)-$ edges suggest
that the dispersion slices and the $L^2_{\kparv}-$ spectra of the $(2,1)-$ and armchair edges are representative of general $(a_1,b_1)-$ edges.

Finally, we remark that the band structure slices,  displayed in the bottom row of Figure \ref{fig_armchair_21_nofold} suggest (for some class of potentials) that the spectral no-fold condition holds for sufficiently large $\eps$. A  rigorous treatment (in preparation) can be provided via semi-classical asymptotic  analysis of the {\it tight-binding} limit. 

\ack
The authors wish to thank I. Aleiner, L. Lu, A. Millis, M. Rechtsman and M. Solja\v{c}i\'c for stimulating discussions.
This work was supported in part by the NSF: DMS-1265524 (CLF), DMS-1412560,  DGE-1069420 (JPL-T, MIW); and the Simons Foundation: \#376319 (MIW).

\section*{References}


\bibliographystyle{iopart-num}
\bibliography{2D-materials}

\end{document}